\documentclass[10pt,a4paper]{article}

\usepackage[
    top=20mm,
    bottom=20mm,
    left=18mm,
    right=18mm
]{geometry}

\usepackage{graphicx}
\usepackage{multicol}
\usepackage{titlesec}
\usepackage{fancyhdr}
\usepackage{setspace}
\usepackage{array}
\usepackage{booktabs}
\usepackage{tikz}
\usepackage{changepage}
\usepackage{caption}
\usepackage{fontspec}
\usepackage{amsmath}
\usepackage{lipsum}
\usepackage{varwidth}
\usepackage{adjustbox}

\usepackage[english]{babel}

\usepackage{url}
\usepackage{amsfonts}
\usepackage{nicefrac}
\usepackage{microtype}
\usepackage{multirow}
\usepackage{tabularx}
\usepackage{lineno}
\usepackage[table,xcdraw,dvipsnames]{xcolor}

\newcolumntype{Y}{>{\raggedright\arraybackslash}X}

\usepackage{siunitx}

\usepackage[most]{tcolorbox}
\tcbuselibrary{breakable}

\definecolor{HeaderGray}{gray}{0.90}
\definecolor{OurModelLight}{gray}{0.97}
\definecolor{OurModelDark}{gray}{0.93}
\definecolor{lavender}{RGB}{216,210,245}
\definecolor{lightlavender}{RGB}{240,242,255}
\definecolor{bestblue}{RGB}{54,85,228}

\definecolor{NRboxbg}{RGB}{245,245,245}
\definecolor{NRboxframe}{RGB}{200,200,200}

\definecolor{headerblue}{RGB}{0,51,102}
\definecolor{rowblue}{RGB}{240,245,255}
\definecolor{tether-gray}{HTML}{F2F1EF}

\usepackage{booktabs,longtable}
\usepackage[table]{xcolor}
\usepackage{multirow}
\usepackage{array}
\usepackage[table]{xcolor} 
\usepackage{siunitx} 

\definecolor{tableheader}{RGB}{240, 242, 245}
\definecolor{tablerow}{RGB}{250, 250, 252}

\newcommand{\nrois}{379}
\newcommand{\model}{BRAID-fMRI }
\newcommand{\clipmodel}{CLIP ViT-B/32}

\usepackage{hyperref}

\hypersetup{
    colorlinks=true,
    linkcolor=bestblue,
    citecolor=bestblue,
    filecolor=magenta,
    urlcolor=cyan
}

\newfontfamily\sharpfont[
    Path=./,
    BoldFont=SharpGrotesk-Medium20.otf,
    ItalicFont=SharpGrotesk-BookItalic20.otf,
    BoldItalicFont=SharpGrotesk-BoldItalic20.otf
]{SharpGrotesk-Book20.otf}

\DeclareCaptionFont{sharp}{\sharpfont}

\titleformat{\section}
    {\sharpfont\large\bfseries}
    {\thesection}
    {1em}
    {}

\titleformat{\subsection}
    {\sharpfont\normalsize\bfseries}
    {\thesubsection}
    {1em}
    {}

\titleformat{\subsubsection}
    {\sharpfont\normalsize\bfseries}
    {\thesubsubsection}
    {1em}
    {}

\titleformat{\paragraph}[runin]
    {\sharpfont\normalsize\bfseries}
    {\theparagraph}
    {1em}
    {}

\newcommand{\captiontitle}[1]{{\sharpfont\bfseries #1}}

\newtcolorbox{reviewbox}[2][]{%
    enhanced,
    breakable,
    colback=tether-gray,
    colframe=tether-gray,
    boxrule=0pt,
    left=15pt,
    right=15pt,
    top=15pt,
    bottom=15pt,
    fonttitle=\sharpfont\bfseries,
    coltitle=black,
    title=#2,
    #1
}

\renewtcolorbox[
  auto counter
]{reviewbox}[2][]{%
  title={Box~\thetcbcounter\ | #2},
  label={#1},
  breakable,
  colback=tether-gray,
  colframe=tether-gray,
  boxrule=0pt,
  arc=5pt,
  outer arc=5pt,
  left=15pt,
  right=15pt,
  top=15pt,
  bottom=15pt,
  toptitle=6pt,
  bottomtitle=4pt,
  coltitle=black,
  fonttitle=\sharpfont\bfseries
}

\definecolor{tableheader}{RGB}{240, 242, 245}
\definecolor{tablerow}{RGB}{250, 250, 252}

\begin{document}


\begin{flushleft}
\includegraphics[height=20pt]{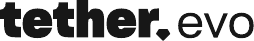}
\end{flushleft}

\vspace{10pt}


\begin{tcolorbox}[
    colback=tether-gray,
    colframe=tether-gray,
    boxrule=0pt,
    arc=5pt,
    left=15pt,
    right=15pt,
    top=15pt,
    bottom=15pt,
    width=\textwidth
]


{\sharpfont\huge\bfseries
Many Brains, One Geometry: A Shared Visual-Semantic Space for Cross-Dataset fMRI Decoding
}

\vspace{5pt}


{\small\bfseries
Moein Khajehnejad\textsuperscript{1,2,}\footnote{The authors contributed equally to this manuscript.},
Michelangelo Tronti\textsuperscript{3,\textcolor{blue}{$a$}}
Forough Habibollahi\textsuperscript{1},
Tommaso Boccato\textsuperscript{1},
Matteo Ferrante\textsuperscript{1,3},
Nicola Toschi\textsuperscript{3}

}

\vspace{2.5pt}

{\small
\textsuperscript{1}Tether Evo, 
\textsuperscript{2}Turner Institute for Brain and Mental Health, Monash University,
\textsuperscript{3}The University of Rome Tor Vergata
}

\vspace{15pt}


{\footnotesize
Visual decoding from fMRI is typically siloed by participant and experiment, obscuring whether heterogeneous neural measurements can be organized within a common computational geometry. Here we introduce \textbf{BRAID-fMRI} (\textbf{B}rain \textbf{R}epresentation \textbf{A}lignment across
\textbf{I}ndividuals and \textbf{D}atasets), a shared CLIP-supervised decoding framework. BRAID-fMRI uses a single ROI-wise Transformer with optional
participant conditioning across eight visual-fMRI datasets comprising 93 dataset-specific participant entries, 430,007 single-trial responses and
162,839 unique stimuli. Regional brain activity is aligned with 512-dimensional CLIP ViT-B/32 representations using a multi-positive contrastive objective that treats repeated stimuli across participants and datasets as positives. BRAID-fMRI supports retrieval across seven evaluation
datasets. On eight matched participant entries, it achieves $35.0\pm11.1\%$ Top-10 accuracy, exceeding the observed mean accuracy of the two evaluated baselines---the MindEye-style pooled-CLIP decoder
($27.1\pm5.1\%$) and ridge regression ($21.1\pm9.1\%$)---and attaining the highest observed accuracy for seven of eight entries. In separately trained participant-agnostic models, expanding the source pool increased target-dataset holdout accuracy by up to $92.1\%$ relative to the initial source-training condition. The learned space preserves graded semantic structure, while ablations and saliency highlight ventral and early visual cortex and
category-specific motion and attentional systems. These results support scalable cross-dataset decoding into a common CLIP-aligned space, with model sensitivity concentrated in ventral and early visual inputs.

}

\end{tcolorbox}

\vspace{10pt}




\section*{Introduction}

Modern visual neuroscience faces a paradox: fMRI datasets are becoming larger
and visual decoders increasingly powerful, yet the resulting models remain
largely confined to the participants and experiments on which they were
trained. Individual studies differ in participants, scanners, acquisition
protocols, experimental tasks, stimulus distributions and sampling density.
Consequently, visual decoders are commonly developed for one participant or
one dataset at a time, yielding highly optimized but fragmented models that are
difficult to compare, scale or transfer across studies. This fragmentation
leaves a fundamental question unresolved: \emph{can heterogeneous neural
measurements support a common visual representational geometry that transfers
to participants and experiments never encountered during training?}
A related challenge is to determine whether such commonality reflects
structure preserved in the neural representations themselves, rather than
simply their alignment to the same pretrained visual target space.

Visual decoding from fMRI has progressed substantially over the past two
decades. Early encoding-based approaches demonstrated that activity in visual
cortex could identify previously unseen natural images
\cite{kay2008identifying}, while subsequent work used deep neural
representations and generative priors to reconstruct increasingly complex
visual content from brain activity
\cite{shen2019deep,takagi2023highresolution}. More recently, large
vision--language models such as CLIP \cite{radford2021learning} have provided
a semantically structured target space in which neural responses can be
aligned with natural images, enabling high-accuracy retrieval and
reconstruction \cite{scotti2023mindeye,scotti2024mindeye2,zhang2025hi}. Voxelwise encoding studies additionally show that CLIP features predict held-out responses in high-level visual cortex \cite{wang2023better}, supporting their use as a shared visual target without establishing that their distances reproduce native cortical geometry. Large, densely
sampled resources such as the Natural Scenes Dataset have been central to this
progress \cite{allen2022massive}. However, much of the strongest performance has been obtained by fitting or adapting mappings to extensively sampled participants within a single experimental setting. Recent approaches have begun to share information across participants. MindEye2, for example, pretrains across participants before adapting to a new individual \cite{scotti2024mindeye2}, but such transfer has largely involved participant-specific adaptation within a common dataset and acquisition regime.
Other recent work has extended this direction across datasets:
NeuroFusion jointly learns across datasets and also evaluates leave-one-dataset-out transfer with target-specific alignment-layer adaptation \cite{kashif2026neurofusion}. Brain-IT shares functional voxel-cluster processing across NSD participants and adapts target voxel embeddings for a new participant \cite{beliy2026brain}. Zero-shot participant transfer has been demonstrated within NSD by ZEBRA \cite{wang2025zebra} and across participants held out from jointly represented datasets by PictorialCortex \cite{huo2026pictorial}. TRIBE~v2 synthetic-only decoders additionally retrieve images from datasets excluded from the response generator's pretraining \cite{benchetrit2026boosting}.

A complementary line of work has shown that neural representations themselves contain structure shared across individuals. Hyperalignment identifies high-dimensional transformations that place participant-specific cortical responses into a common representational space \cite{haxby2011common}, while shared-response models explicitly factor multi-participant fMRI data into individual mappings and a shared latent response \cite{chen2015sharedresponse}. Representational similarity analysis provides a complementary framework for comparing the geometry of neural and computational representations independently of their native coordinate systems \cite{kriegeskorte2008rsa,kriegeskorte2013representational}.
Recent work has gone further, showing that independently learned
participant-specific fMRI spaces within NSD can exhibit approximately isometric geometry and be related through unsupervised orthogonal transformations \cite{marcos2026platonic}. Those spaces were learned from repeated neural measurements without visual-model targets; here, geometry is evaluated after supervised alignment to CLIP.
Together, these studies establish that common structure can be recovered across people under coordinated experimental conditions. Cross-dataset learning poses a more stringent problem: inter-individual variability is compounded by differences in scanner hardware, spatial and temporal resolution, experimental design, stimulus statistics, and the amount of data available for
each participant. It therefore remains unclear whether a common
relational geometry persists when representations are learned jointly across independently acquired datasets, and whether that geometry can support direct transfer to participants and experiments absent from training.

MOSAIC makes it possible to test this question at a substantially broader scale \cite{lahner2025mosaic}. The resource harmonizes eight independently collected visual-fMRI datasets: the Natural Scenes Dataset (NSD), THINGS-fMRI (THINGS), Deeprecon, BOLD5000, the BOLD Moments Dataset (BMD), Generic Object Decoding (GOD), the Human Actions Dataset (HAD) and the Natural Object Dataset (NOD). MOSAIC comprises 93 dataset-specific participant entries from 77 individuals, 430,007 single-trial fMRI--stimulus pairs and 162,839 unique stimuli. Data are uniformly reprocessed with fMRIPrep, registered to the fsLR32k cortical surface and converted to single-trial beta estimates with GLMsingle, with curated splits preventing duplicate or perceptually similar train--test stimuli. These datasets sample different portions of both the human population and the visual world. MOSAIC itself establishes a framework for aggregating these datasets and demonstrates the value of multi-dataset training for brain encoding. Here we use this harmonized resource to address a distinct decoding question: whether neural responses from all eight datasets can be mapped by one shared model into the same visual-semantic coordinate system, whether that mapping transfers without target-participant calibration, and whether its internal brain-space geometry is shared across datasets. Rather than treating this heterogeneity solely as nuisance variation to be removed, we ask whether it can provide complementary training signal from which a representation transferable across both datasets and individuals can emerge.

We address this question with a single ROI-wise brain-to-CLIP model trained jointly across all eight datasets. Anatomically corresponding regions of interest are mapped into a common token space using ROI-specific encoders, augmented with learned ROI embeddings and, when used, participant embeddings, and integrated by a Transformer shared across every participant and dataset
\cite{vaswani2017attention} (Fig.~\ref{fig:schematic}). Attention pooling produces a brain representation projected into the frozen 512-dimensional image-embedding space of CLIP ViT-B/32 \cite{radford2021learning}. A symmetric multi-positive contrastive objective allows repeated observations of the same stimulus, including repetitions across participants or datasets, to reinforce a common representation rather than act as false negatives. Learned participant embeddings provide a lightweight mechanism for accommodating participant-specific variation when identity is available, paralleling the participant alignment or adaptation used in previous cross-participant decoders \cite{ferrante2024through,scotti2024mindeye2}. Crucially, participant conditioning is optional; disabling participant embeddings provides a participant-agnostic shared encoder that can be evaluated directly on participants and datasets absent from training. This design therefore separates two questions that are often conflated: whether participant-specific conditioning can accommodate individual variation when identity is known, and whether the shared representation itself is sufficiently invariant to operate without such conditioning.

We find that a single model supports stimulus retrieval across seven evaluation datasets and, across eight matched participants, performs best for seven relative to Ridge regression \cite{hoerl1970ridge} and a MindEye-style retrieval baseline. Following Benchetrit et al.~\cite{benchetrit2026boosting}, this MindEye-style comparison predicts a single pooled CLIP embedding for retrieval, without the diffusion prior or image reconstruction used in the full MindEye framework \cite{scotti2023mindeye}. Separately trained participant-agnostic models also support target-dataset holdout retrieval without fitting participant embeddings. Expanding the source pretraining pool was associated with relative gains of up to 92.1\% in target-dataset holdout accuracy; this comparison jointly changes dataset diversity and the amount of training data. Within the CLIP-supervised prediction space, representations exhibit a graded hierarchy from stimulus identity to semantic relatedness, and category-level geometry shows correspondence among the evaluable dataset pairs.
Functional-network ablations and ROI-level saliency further reveal a common dependence on early and ventral visual systems alongside category-dependent contributions from motion and attentional networks. Our contribution is the joint evaluation of one anatomically indexed decoder across heterogeneous visual-fMRI datasets: matched-participant retrieval, target-dataset holdout transfer without fitted participant embeddings, and characterization of semantic geometry and anatomical sensitivity within the CLIP-supervised prediction space.

\section*{Results}

\subsection*{A shared cross-dataset encoder combines anatomical correspondence and optional participant conditioning}

\begin{figure}[!h]
    \centering
    \includegraphics[width=\textwidth]{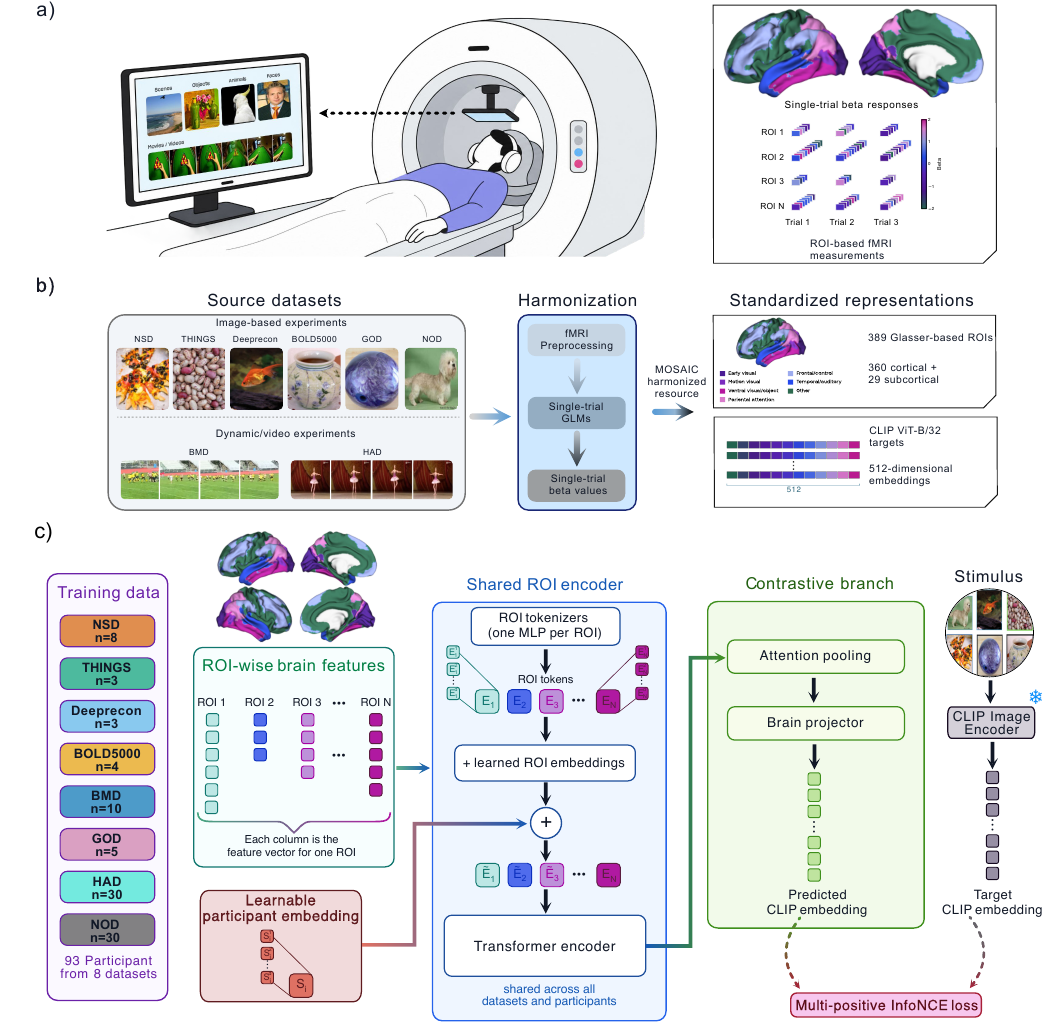}
    \caption{
\captiontitle{Cross-dataset visual-fMRI harmonization and the \model model.}
\textbf{a,} Visual stimuli are presented during fMRI acquisition, and
single-trial responses are estimated to obtain stimulus-specific brain activity
patterns.
\textbf{b,} Eight visual-fMRI datasets spanning static-image and dynamic-video
experiments are brought into a common processing framework. fMRI responses are
represented using 360 HCP-MMP1.0 cortical parcels and 19 additional regions, while
stimuli are represented in a common 512-dimensional CLIP ViT-B/32 image space.
\textbf{c,} ROI-wise responses from participants in eight visual-fMRI datasets
are mapped to ROI tokens and augmented with learned ROI embeddings and, in
participant-conditioned analyses, learned participant embeddings. A Transformer shared
across datasets and participants contextualizes the tokens, which are
attention-pooled and projected into the 512-dimensional image space of a
frozen \clipmodel{} encoder. Participant embeddings are omitted in the zero-shot
transfer experiments. A multi-positive contrastive objective aligns neural
responses with their corresponding stimuli while treating repeated
observations of the same stimulus as positives.
}
    \label{fig:schematic}
\end{figure}

We trained BRAID-fMRI, a single model using the naturalistic MOSAIC training split, pooling
single-trial responses across all eight source datasets and all dataset-specific
participant entries. Brain activity was represented in a common atlas space using
\nrois{} cortical and subcortical regions of interest (ROIs). Each ROI contains a
variable number of cortical vertices or subcortical voxels; therefore, a separate
two-layer tokenizer mapped each ROI vector into a common 512-dimensional token.
A learned ROI identity embedding encoded anatomical correspondence, whereas a
learned participant embedding was broadcast across all ROI tokens from the same
dataset-specific participant. The resulting tokens were processed by a four-layer
Transformer whose parameters were shared across all datasets and participants.
Attention pooling and a projection head produced a 512-dimensional prediction in
the image space of the frozen \clipmodel{} encoder (Fig.~\ref{fig:schematic}).

This construction provides separate parameter components for ROI identity and participant conditioning. ROI-specific tokenizers
accommodate differences in regional dimensionality, while participant embeddings
provide a lightweight participant-dependent shift of the common token space.
The shared Transformer must nevertheless learn interactions among anatomically
corresponding regions that are useful across the complete training collection.
Repeated presentations and stimuli shared between participants or datasets were
handled by a multi-positive InfoNCE objective, preventing valid stimulus matches
from being incorrectly treated as negatives. We use participant conditioning when
evaluating held-out stimuli from participant entries represented during training;
in the zero-shot experiments below, participant embeddings are disabled entirely to
isolate transfer supported by the shared encoder.

\subsection*{One model retrieves visual stimuli across datasets and participants}

\begin{figure}[!h]
    \centering
    \includegraphics[width=0.77\textwidth]{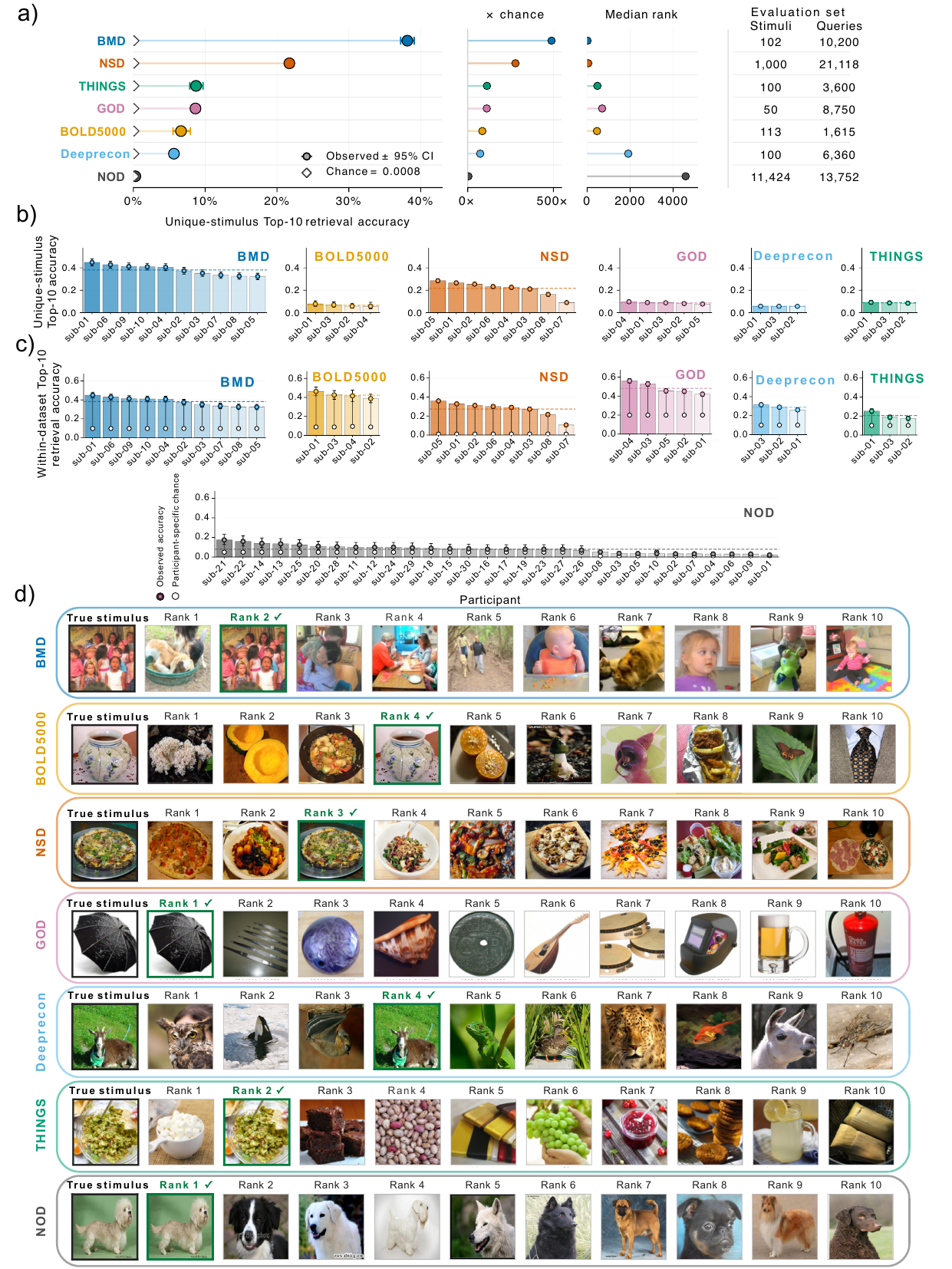}
    \caption{\captiontitle{Cross-dataset and within-dataset stimulus retrieval.}
    \textbf{a,} Dataset-level unique-stimulus Top-10 retrieval using a global candidate pool
comprising all unique validation stimuli across the seven evaluation datasets.
Filled circles show observed accuracy with query-level 95\% Wilson confidence
intervals, which do not account for within-participant or within-session
dependence;
open diamonds denote global chance. Adjacent summaries show performance relative
to chance ($\times$ chance), median rank of the correct stimulus (lower is
better), and the numbers of candidate stimuli and retrieval queries.
\textbf{b,} Participant-level Top-10 retrieval using the same global candidate
pool. Participants are ordered by accuracy within each dataset; filled circles
and error bars show observed accuracy and 95\% Wilson confidence intervals, and
dashed lines indicate dataset means. NOD is omitted for visual clarity because
of its larger participant sample.
\textbf{c,} Participant-level retrieval with the candidate pool restricted to
unique validation stimuli from the query participant's dataset. Filled circles
and error bars show observed accuracy and 95\% Wilson confidence intervals;
open circles denote participant-specific chance levels and vertical connectors
show the difference from chance. Dashed lines indicate dataset means. NOD is
shown separately in a wider lower panel.
\textbf{d,} Each row shows the true stimulus followed by the ten highest-ranked
candidates based on cosine similarity between the predicted brain embedding and
frozen CLIP ViT-B/32 image embeddings. Green borders and check marks identify
the true stimulus. Examples span multiple datasets; high-ranking non-matches
often share object category, scene context, shape or texture with the target,
illustrating the semantic structure of the learned retrieval space. HAD dataset is not presented since it is not included in the retrieval evaluation.}
    \label{fig:retrieval}
\end{figure}

We first asked whether a single model trained jointly across heterogeneous
datasets and participants could identify held-out visual stimuli from fMRI.
For each query, the predicted brain embedding was compared by cosine
similarity with precomputed CLIP ViT-B/32 image embeddings, and unique
candidate stimuli were ranked accordingly. In the most stringent evaluation,
the candidate pool comprised all unique validation stimuli across the seven
evaluation datasets. Dataset-level unique-stimulus Top-10 accuracy reached
38.1\% for BMD, 21.7\% for NSD, 8.7\% for THINGS, 8.6\% for GOD, 6.6\% for
BOLD5000, 5.6\% for Deeprecon and 0.3\% for NOD, compared with a global chance
level of 0.08\% (Fig.~\ref{fig:retrieval}a). Observed accuracy exceeded the global random-ranking reference across datasets, despite differences in their stimulus distributions and numbers of evaluation queries. Participant-level performance also varied within datasets,
indicating heterogeneity in decoding difficulty across individuals and
experimental settings (Fig.~\ref{fig:retrieval}b).

We next restricted the candidate pool to unique validation stimuli from the
query participant's dataset. This dataset-restricted evaluation generally
increased retrieval accuracy, particularly for GOD, BOLD5000, Deeprecon and
THINGS (Fig.~\ref{fig:retrieval}c). Because the number of candidate stimuli
differed across datasets, the corresponding Top-10 chance levels also varied.
The improvement relative to the global setting indicates that part of the
retrieval challenge arises from discriminating among stimuli drawn from
multiple experiments and visual domains. Accuracy above global random-ranking chance does not by itself exclude dataset-specific cues: a predictor that identifies the source dataset can already restrict the plausible candidate set. Stimulus discrimination therefore also needs to be evaluated against the corresponding restricted-candidate reference.

We then compared \model with Ridge regression and a MindEye-style pooled-CLIP decoder using the
eight participants for whom matched results were available in BOLD5000 and NSD. Across this subset, \model achieved a mean Top-10 accuracy of 35.0\%, compared with 27.1\% for MindEye and 21.1\% for Ridge regression (Table~\ref{tab:baselines}). It achieved the highest participant-level accuracy in seven of the eight comparisons and the highest dataset-level mean for both BOLD5000 and NSD. 
Thus, jointly learning from heterogeneous datasets did not require sacrificing participant-level performance and, under the matched evaluation considered here, generally outperformed established participant-specific baselines. A broader comparison against Ridge regression across 63 dataset--participant
entries is provided in Supplementary Table~\ref{tab:baselines-expanded}; \model achieved a higher overall mean ($22.2\%$ vs.\ $13.1\%$) and a
higher dataset-level mean in six of seven datasets.

Representative retrieval examples further illustrated the structure of the
model's predictions (Fig.~\ref{fig:retrieval}d). When the exact target was not ranked first, the highest-ranked
candidates frequently preserved salient semantic or perceptual properties of
the target, including object category, scene context, shape and texture.
These selected examples illustrate semantically or perceptually related non-matches, suggesting that retrieval errors were typically structured within the learned visual-semantic space.

\begin{table}[h]
\centering
\small
\begin{minipage}{0.82\textwidth} 
\centering

\definecolor{headbg}{gray}{0.92}
\definecolor{meanbg}{gray}{0.96}
\definecolor{bestbg}{gray}{0.88}

\renewcommand{\arraystretch}{1.05}
\setlength{\tabcolsep}{4.5pt}

\caption{\textbf{Participant-matched unique-stimulus Top-10 retrieval accuracy.}}
\label{tab:baselines}
\begin{tabular}{llccc}
\toprule
\rowcolor{headbg}
\textbf{Dataset} & \textbf{Participant} & \textbf{Ridge regression} & \textbf{MindEye-style} & \textbf{\model} \\
\midrule
\multirow{5}{*}{BOLD5000}
 & sub-01 & 26.8 & 34.0 & \textbf{46.3} \\
 & sub-02 & 22.6 & 34.0 & \textbf{38.8} \\
 & sub-03 & 23.3 & 30.0 & \textbf{42.6} \\
 & sub-04 & 35.9 & 26.0 & \textbf{41.6} \\
 \rowcolor{meanbg}
 & Mean $\pm$ s.d. & 27.2 $\pm$ 6.1 & 31.0 $\pm$ 3.8 & \textbf{42.3 $\pm$ 3.1} \\
\midrule
\multirow{5}{*}{NSD}
 & sub-01 & 18.6 & 23.0 & \textbf{32.8} \\
 & sub-02 & 17.1 & 22.0 & \textbf{31.1} \\
 & sub-03 & 20.3 & 27.0 & \textbf{35.9} \\
 & sub-04 & 3.9 & \textbf{21.0} & 10.6 \\
 \rowcolor{meanbg}
 & Mean $\pm$ s.d. & 15.0 $\pm$ 7.5 & 23.3 $\pm$ 2.6 & \textbf{27.6 $\pm$ 11.5} \\
\midrule
\rowcolor{meanbg}
Overall & Mean $\pm$ s.d. & 21.1 $\pm$ 9.1 & 27.1 $\pm$ 5.1 & \textbf{35.0 $\pm$ 11.1} \\
\rowcolor{bestbg}
 & Best-performing participants & 0/8 & 1/8 & \textbf{7/8} \\
\bottomrule
\end{tabular}

\vspace{4pt}
\raggedright\footnotesize
Values are percentages. Comparisons include the eight BOLD5000 and NSD
participants available for all three methods. Following the retrieval-only MindEye-style approach of
  Benchetrit et al.~\cite{benchetrit2026boosting}, our baseline predicts a single pooled
  image embedding for contrastive retrieval. Unlike the full MindEye framework
  \cite{scotti2023mindeye}, it does not model the distribution of CLIP image
  embeddings with a diffusion prior or perform image reconstruction. 
\end{minipage}
\end{table}

\subsection*{The participant-agnostic encoder supports transfer in target-dataset holdout evaluations}

We next asked whether cross-dataset transfer depended on learned participant
identities. For these experiments, participant embeddings were disabled throughout
training and evaluation. Each target dataset---including all of its participants
and stimuli---was excluded from training, while the remaining source pool was
expanded cumulatively by adding datasets and their selected participants.
Thus, no target-participant embedding or other target-specific decoder parameter was fitted.

In this participant-agnostic setting, the final curriculum condition exceeded
the initial condition for all seven targets (Fig.~\ref{fig:clusters}g). Relative
gains from the first to the final step were $92.1\%$ for NSD, $69.0\%$ for BMD,
$60.9\%$ for BOLD5000, $32.0\%$ for NOD, $25.3\%$ for THINGS, $11.7\%$ for
Deeprecon and $10.9\%$ for GOD. Individual curriculum steps were not uniformly
beneficial, indicating that source datasets differ in their relationship to a
given target. Nevertheless, the positive final gain for every target shows that
the shared encoder captures cross-participant and cross-dataset structure that can
transfer without participant-specific conditioning, with a positive final relative gain in each of the seven reported target comparisons.

\subsection*{Retrieval performance follows visual-semantic organization}

\begin{figure}[!h]
    \centering
    \includegraphics[width=0.85\textwidth]{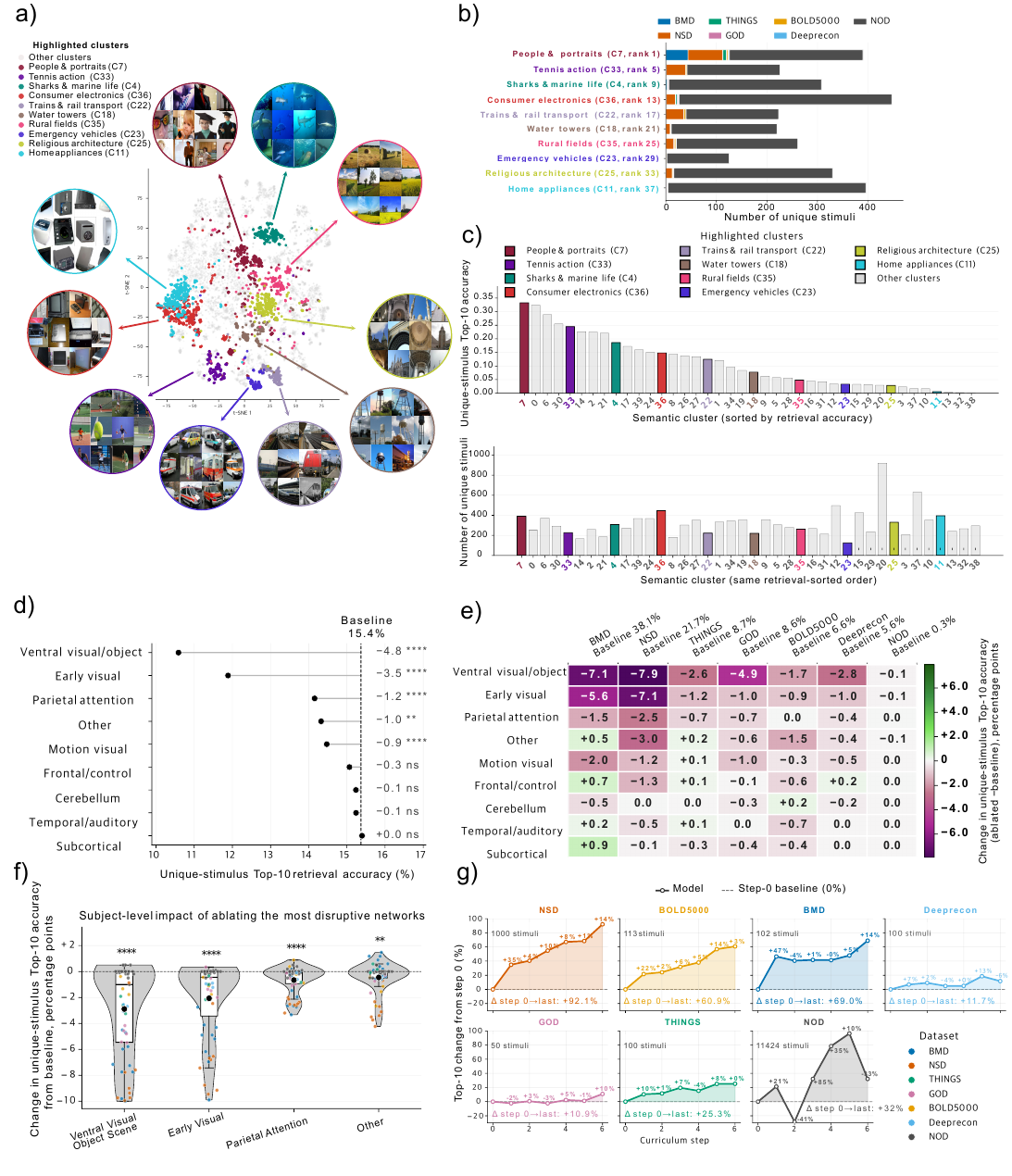}
    \caption{\captiontitle{Semantic organization, network contributions and zero-shot cross-dataset transfer.}
    \textbf{a,} t-SNE visualization of 40 $k$-means clusters obtained from
    normalized CLIP embeddings of unique validation stimuli. Ten clusters used
    for subsequent analyses are highlighted, with representative stimuli shown.
    \textbf{b,} Dataset composition of the selected clusters. Stacked bars
show the number of unique stimuli contributed by each dataset. Cluster labels (C0--C39) denote arbitrary $k$-means cluster identifiers and
are unrelated to retrieval rank; ranks indicate ordering by Top-10 retrieval
accuracy. \textbf{c,} Top:
    unique-stimulus Top-10 accuracy for all clusters, ordered by performance;
    bottom: number of unique stimuli in the same ordering. Every fourth cluster
    in the ranked sequence was retained, yielding ten clusters spanning the full
    performance range.
     \textbf{d,} Global unique-stimulus Top-10 accuracy after masking
    each functional or anatomical ROI group. The dashed line marks the unablated
    baseline; annotations give percentage-point changes. \textbf{e,}
    Dataset-specific ablation effects. \textbf{f,} Participant-level effects for
    the four most disruptive groups. Points denote participants, violins show
    distributions, boxes show medians and interquartile ranges, and black circles
    show means. Asterisks denote one-sided Wilcoxon signed-rank tests of retrieval
    impairment, with Benjamini--Hochberg correction across nine groups.
    \textbf{g,} Zero-shot participant and dataset transfer with participant embeddings
disabled throughout training and evaluation. For each target, the entire
dataset and all of its participants remain held out while source datasets and
participants are cumulatively added to pretraining. Values are percentage
changes relative to step 0; annotations report the net change from the first
to the final step.}
    \label{fig:clusters}
\end{figure}

The large, heterogeneous candidate pool enabled us to ask whether retrieval
performance was structured by stimulus semantics. We clustered unique validation
stimuli by applying $k$-means ($k=40$) to their $\ell_2$-normalized CLIP
embeddings, using t-SNE only to visualize the resulting organization. The clusters
formed recognizable semantic groups, including people and portraits, tennis
actions, marine life, consumer electronics, trains, rural scenes and architecture
(Fig.~\ref{fig:clusters}a). Descriptive cluster names were assigned manually
after inspecting representative stimuli from each cluster and played no role in
the clustering procedure. Most selected clusters contained stimuli contributed
by multiple datasets, although NOD supplied a large fraction of many categories
because of its broad stimulus set (Fig.~\ref{fig:clusters}b). Note that cluster IDs (C0--C39) are arbitrary $k$-means labels and do not
encode any particular ordering. A complete atlas of all 40 clusters, together with their retrieval performance,
size and cross-dataset composition, is shown in Supplementary
Fig.~\ref{fig:supp_all_clusters}.

Retrieval differed markedly across clusters. People and portraits achieved the
highest Top-10 accuracy, followed by several coherent categories including tennis
actions and marine life, whereas home appliances and some architectural categories
were among the most difficult (Fig.~\ref{fig:clusters}c). High- and low-performing groups occurred across a range of stimulus counts and dataset compositions (Supplementary Fig.~\ref{fig:supp_all_clusters}). To avoid selecting categories after inspecting their content, we ranked all 40 clusters by retrieval accuracy and retained every fourth cluster, yielding ten categories at ranks 1, 5, 9, \ldots, 37. This performance-stratified sampling scheme spans the observed accuracy range and provides an illustrative subset for subsequent geometric and neuroanatomical analyses.

\subsection*{A shared visual backbone supports retrieval across datasets}

We next examined the trained model's sensitivity to masking inputs from different anatomical and functional groups.
Atlas ROIs were assigned to nine functional or anatomical groups, and all
tokens from one group were masked at inference. The unablated pooled Top-10
accuracy was 15.4\%. Masking ventral visual/object regions produced the largest
reduction, decreasing accuracy by 4.8 percentage points, followed by early
visual cortex ($-3.5$ points), parietal attention regions ($-1.2$), the
heterogeneous ``Other'' group ($-1.0$) and motion-selective visual regions
($-0.9$; Fig.~\ref{fig:clusters}d). These reductions survived
Benjamini--Hochberg correction across the nine ablations, whereas
frontal/control, cerebellar, temporal/auditory and subcortical ablations had
small pooled effects. The dominance of ventral and early visual systems
indicates that cross-dataset alignment remains anchored in core visual
representations rather than being distributed uniformly across the brain.

The same hierarchy was broadly visible at the participant level, but effect
sizes varied across datasets (Fig.~\ref{fig:clusters}e,f). Ventral and early
visual ablations were especially disruptive for BMD and NSD, whereas the
contribution of other systems varied more strongly across datasets. The shared
representation therefore does not depend on a single invariant subset of
regions. Instead, it combines a common early and ventral visual backbone with
content- and dataset-dependent contributions from attention, motion and other
systems. The complete cluster-by-network ablation landscape across all 40
semantic clusters is shown in Supplementary
Fig.~\ref{fig:supp_cluster_networks}a.

\subsection*{The learned space preserves graded semantics and category-specific cortical structure}

\begin{figure}[!h]
    \centering
    \includegraphics[width=\textwidth]{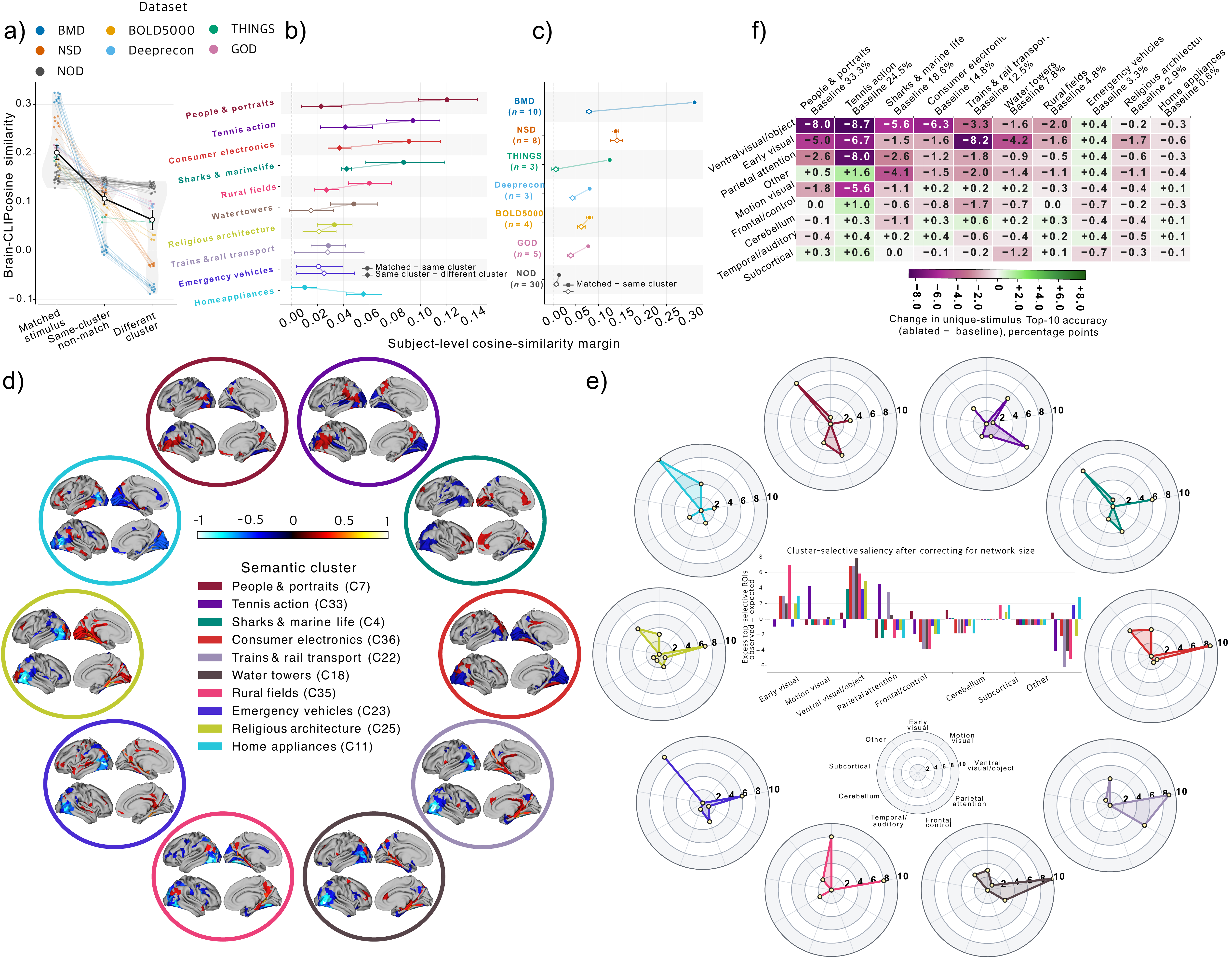}
    \caption{\captiontitle{CLIP-space semantic organization and category-associated attribution
patterns.} \textbf{a,} Brain--CLIP cosine similarity for the
    matched stimulus, a non-matching stimulus from the same semantic cluster and
    a stimulus from a different cluster. Thin lines show participant-level or
    dataset-level trajectories, and summary markers show the group mean with
    uncertainty. \textbf{b,} Participant-level identity and semantic margins for the
    ten representative clusters, defined as matched minus same-cluster similarity
    and same-cluster minus different-cluster similarity, respectively.
    \textbf{c,} The same margins summarized by dataset.
    \textbf{d,} Cluster-associated ROI beta patterns projected
    onto the HCP-MMP1.0 cortical surface for ten representative semantic
    clusters. Colours indicate normalized cluster-associated beta values.
    \textbf{e,} Network composition of positively cluster-selective saliency.
    Radar plots show observed counts among up to 20 positively selective ROIs in
    each network for individual clusters; the central bar plot shows observed
    minus expected counts after correcting for network size. Positive values
    indicate enrichment.  \textbf{f,}
    Cluster-specific change in unique-stimulus Top-10 accuracy after ablating each
    functional or anatomical network. Values are ablated minus baseline accuracy
    in percentage points; negative values indicate that the ablated network
    supported retrieval.}
    \label{fig:saliency}
\end{figure}

A successful retrieval model could in principle separate individual stimuli without
preserving relationships among them. We therefore compared the cosine similarity
between each predicted brain embedding and three classes of CLIP targets: its
matched stimulus, a non-matching stimulus from the same semantic cluster and a
stimulus from a different cluster. Across participants, matched pairs were more
similar than same-cluster non-matches by a mean margin of 0.094
($P=6.4\times10^{-12}$). Same-cluster non-matches were, in turn, more similar than
different-cluster stimuli by a mean margin of 0.043
($P=3.9\times10^{-10}$; Fig.~\ref{fig:saliency}a). This ordering was visible
across the seven evaluation datasets and across the representative semantic
clusters, although the magnitude of the margins varied by dataset and category
(Fig.~\ref{fig:saliency}b,c). The model therefore preserves both stimulus identity
and broader semantic neighbourhood structure. The same analysis across all 40 semantic clusters showed substantial variation
in the strength of both margins while preserving the broader hierarchical
pattern (Supplementary Fig.~\ref{fig:supp_cluster_networks}b).

We next examined how this geometry was expressed across cortical systems. For each
query, ROI saliency was defined as the absolute gradient-times-input contribution to
matched brain--CLIP cosine similarity. Cluster selectivity contrasted an ROI's mean
saliency for one semantic cluster with its mean saliency for all other clusters,
standardized by pooled variability. Cluster-associated beta maps were spatially
distributed and differed systematically among categories (Fig.~\ref{fig:saliency}d).
People and portraits, tennis actions, marine life, consumer electronics and
scene-related categories emphasized partially distinct patterns across occipital,
ventral temporal, parietal and motion-sensitive cortex.

To distinguish selectivity from differences in network size, we compared the number of top positively selective ROIs in each network with the count expected from that network's share of all ROIs. Early visual and ventral visual/object systems were enriched across several clusters, whereas the balance of network contributions varied substantially across semantic content (Fig.~\ref{fig:saliency}e; Supplementary
Fig.~\ref{fig:supp_cluster_networks}c). Cluster-specific ablations provided complementary functional evidence (Fig.~\ref{fig:saliency}f). People and portraits were most impaired by ventral visual/object ablation ($-8.0$ percentage points). Tennis actions depended jointly on ventral visual/object ($-8.7$), parietal attention ($-8.0$) and early visual regions ($-6.7$), whereas trains and rail transport were particularly sensitive to early visual ablation ($-8.2$). Thus, the trained decoder shows broadly shared visual-input dependence and category-associated variation in masking effects, although category and dataset composition are not disentangled here.

\subsection*{Shared semantic geometry persists across datasets and retrieval difficulty}

The preceding analyses showed that the model preserves semantic structure relative
to CLIP and recruits partly distinct cortical systems across stimulus categories.
We next asked whether CLIP-supervised brain predictions preserve semantic relationships across datasets and remain informative as retrieval becomes progressively more difficult.

\begin{figure}[!h]
    \centering
    \includegraphics[width=\textwidth]{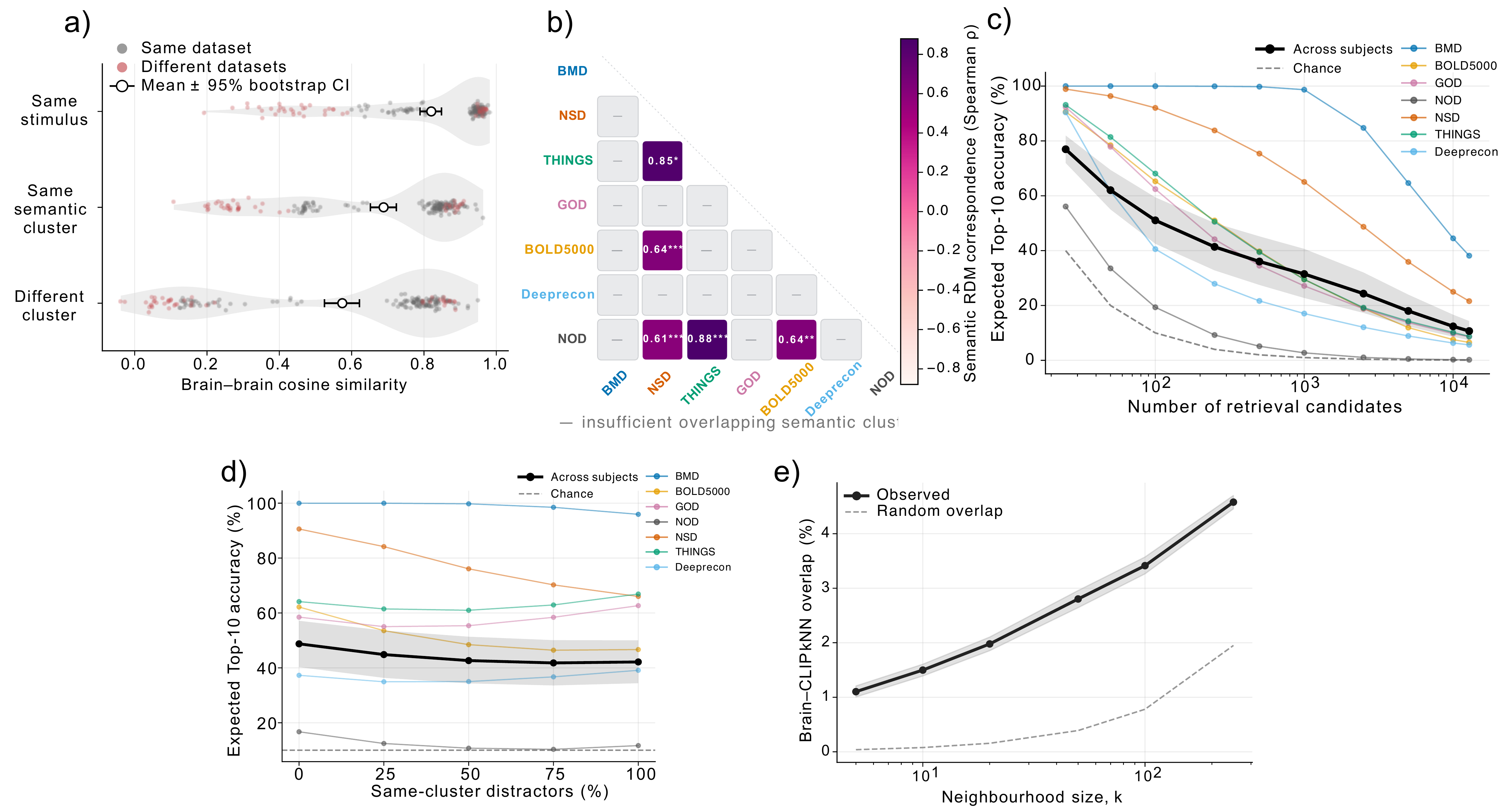}
    \caption{
    \captiontitle{Semantic organization and robustness of the learned brain representation.}
    \textbf{a,} Hierarchical structure within the learned brain-embedding space.
    Brain--brain cosine similarity is compared across participant pairs for the same
    stimulus, different stimuli from the same CLIP-derived semantic cluster, and
    stimuli from different clusters. Points distinguish participant pairs drawn from the  same or different datasets; summary markers show the mean and 95\% bootstrap confidence interval.
    \textbf{b,} Cross-dataset correspondence of semantic representational geometry. Within each dataset, a representational dissimilarity matrix (RDM) was constructed from cosine distances among semantic-cluster centroids. Cells report Spearman correlations between RDMs for dataset pairs with at least five shared semantic clusters. Asterisks denote significance after Benjamini--Hochberg correction of Mantel-style permutation tests; dashes indicate insufficient cluster overlap.
    \textbf{c,} Expected retrieval performance as the candidate search space increases. For each query, exact expected Top-10 accuracy was derived from its full-pool rank using the hypergeometric distribution. Coloured curves show dataset means, the black curve shows the mean across participants, and shading denotes the 95\% bootstrap
    confidence interval. The dashed line indicates random Top-10 chance,
    $\min(10/N,1)$, for a candidate pool of size $N$.
    \textbf{d,} Retrieval among semantic hard negatives. The fraction of distractors drawn from the same semantic cluster as the target was increased from 0 to 100\% while candidate-pool size remained fixed. Expected Top-10 accuracy was computed analytically from the numbers of same- and different-cluster candidates outranking each target. Coloured curves show dataset means, the black curve shows the mean across participants, shading denotes the 95\% bootstrap confidence interval and the dashed line indicates random Top-10 chance.
    \textbf{e,} Preservation of local semantic neighbourhoods. Predicted brain
    embeddings were averaged across repeated observations of each unique stimulus. For each neighbourhood size $k$, overlap between the $k$ nearest neighbours in brain and CLIP space was computed across unique stimuli. The black curve shows mean observed overlap with its 95\% bootstrap confidence interval; the dashed curve shows the analytical expectation under random neighbourhood correspondence.
    }
    \label{fig:generalization}
\end{figure}

We first compared pairs of predicted brain embeddings without using CLIP vectors in the pairwise similarity calculation. Across participant pairs, representations of the same stimulus were most
similar, followed by representations of different stimuli from the same semantic
cluster, whereas stimuli from different clusters were least similar
(Fig.~\ref{fig:generalization}a). Participant pairs drawn both within and across
datasets contributed to this graded organization, indicating that the hierarchy
from stimulus identity to broader semantic relatedness is not confined to a
single experimental setting.

We then asked whether the broader arrangement of semantic categories was itself
conserved across datasets. For each dataset, we constructed a representational
dissimilarity matrix from the pairwise cosine distances among semantic-cluster
centroids and compared these geometries between datasets. The five evaluable dataset pairs, involving NSD, THINGS, BOLD5000 and NOD, showed positive correspondence after multiple-comparison correction, with Spearman correlations ranging from $0.61$ to $0.88$
(Fig.~\ref{fig:generalization}b). Thus, datasets with different participants,
stimulus distributions and acquisition protocols nevertheless preserve similar
\emph{relative relationships} among semantic categories. BMD, GOD and Deeprecon lacked sufficient shared clusters for this comparison. The observed correspondences establish relational consistency in the supervised prediction space, without separating target-imposed structure from structure supported by the neural measurements.

We next quantified how retrieval scales as the candidate space expands. Using each
query's rank in the full candidate set, we computed the exact expected Top-10
accuracy for progressively larger candidate pools. Performance declined as the
number of competing stimuli increased, but the across-participant trajectory remained
above random Top-10 chance throughout the tested range
(Fig.~\ref{fig:generalization}c). Dataset-specific trajectories differed
substantially, with some datasets retaining strong retrieval even in large search
spaces. These results show that the learned ranking structure remains informative
as the discrimination problem becomes increasingly demanding.

We additionally tested retrieval with distractors from the target's CLIP-derived semantic cluster rather than arbitrary distractors. We therefore progressively
increased the fraction of candidate distractors drawn from the target's own
CLIP-derived semantic cluster while holding candidate-pool size fixed. Expected
Top-10 accuracy declined only moderately at the population level and remained well
above chance even when all distractors were drawn from the same semantic cluster
(Fig.~\ref{fig:generalization}d). Small non-monotonic changes in individual datasets
are expected because same-cluster membership constrains semantic relatedness but does
not guarantee greater difficulty for every individual query. Retrieval therefore
cannot be explained solely by coarse category identification: the learned
representation retains information that discriminates individual stimuli within
semantic neighbourhoods.

Finally, we asked whether local relationships among unique stimuli were preserved
between brain and CLIP space. Across neighbourhood sizes from $k=5$ to $k=250$,
the overlap between brain-space and CLIP-space nearest neighbours consistently
exceeded the analytical expectation under random neighbourhood correspondence
(Fig.~\ref{fig:generalization}e). Absolute overlap remained modest, indicating
partial preservation rather than replication of CLIP's local topology. Nevertheless, its systematic elevation above chance shows that stimuli occupying nearby locations in CLIP space also tend to remain neighbours in the learned brain representation.

Together, these analyses show that the learned representation supports substantially more than pointwise brain--image alignment. Its CLIP-supervised predictions preserve graded semantic relationships across participant entries and relational correspondence among the evaluable datasets, remains informative under large and semantically challenging candidate spaces, and partially preserves the local organization of the target visual-semantic space.

\section*{Discussion}

By jointly modelling eight visual-fMRI datasets, we show that heterogeneous neural
measurements can be organized within a common visual-semantic representation despite large differences in participants, stimulus distributions and acquisition protocols.
A single model retrieved held-out stimuli across seven evaluation datasets and
outperformed matched participant-specific baselines for seven of eight participants.
More importantly, the shared encoder did not require learned participant identities to generalize: participant-agnostic models supported target-dataset holdout retrieval, with the final source-pool condition improving accuracy by up to $92.1\%$ relative to the initial condition. The learned representation was also structured beyond retrieval accuracy; stimulus identity and semantic relatedness were reflected directly in brain--brain geometry, semantic organization was conserved across datasets, and retrieval remained robust as both the candidate space and prevalence of semantic hard negatives increased. Together, these findings support reusable mappings from heterogeneous visual-fMRI measurements into a shared supervised prediction space.

A central design principle is to provide participant-specific conditioning while sharing the remaining model parameters.
Why retain participant conditioning if the encoder can transfer without it?
Anatomical harmonization does not eliminate inter-individual differences in
functional organization and measurement characteristics, and previous
cross-participant decoders have therefore used explicit participant alignment or
participant-specific adaptation to map individual responses into shared spaces
\cite{ferrante2024through,scotti2024mindeye2}. Here, participant embeddings
provide a considerably lighter mechanism: a single learned vector can modulate
all ROI tokens for a known participant while the ROI tokenizers, Transformer,
pooling and projection remain shared. Their role is therefore to provide
capacity for participant-specific variation, not to define the common
representation itself. The zero-shot results make this distinction explicit:
models trained without participant embeddings still support target-dataset holdout retrieval.

The zero-shot curriculum analysis further indicates that cross-dataset heterogeneity can be a source of transferable information rather than merely a nuisance to be removed. With participant embeddings disabled, every target dataset and all of its participants remained unseen throughout training, while the source pool was progressively expanded. The final, more diverse pretraining condition outperformed the initial condition for all seven targets, although individual curriculum steps were not uniformly beneficial. This non-monotonicity is informative: additional neural data are not interchangeable simply because they increase sample size, and their utility depends on relationships among source and target stimulus distributions, participants and acquisition regimes. Nevertheless, the uniformly positive final gains demonstrate that broader pretraining can expose regularities that transfer directly to unseen
participants and experiments. Controlled scaling experiments that independently manipulate numbers of trials, participants, unique stimuli and datasets will be important for determining which forms of diversity contribute most strongly to this transfer.

The geometry analyses provide complementary evidence that the model learns more than a collection of pointwise brain--image associations. Brain embeddings corresponding to the same stimulus were most similar, embeddings from different stimuli within the same semantic cluster occupied an intermediate regime, and embeddings from unrelated clusters were least similar. Importantly, this ordering persisted for comparisons across datasets. At a broader scale, semantic representational geometries showed substantial correspondence between dataset pairs with sufficient overlapping categories, with RDM correlations ranging from 0.61 to 0.88. These results imply that different experiments can preserve similar \emph{relations} among semantic categories even when their absolute neural measurements differ. The shared space is therefore not simply aligning isolated targets; it is recovering aspects of a common relational organization across the evaluated datasets within the supervised prediction space.

The retrieval stress tests reinforce this interpretation. Accuracy decreased as the candidate pool expanded over several orders of magnitude, as expected, but remained above the corresponding random baseline throughout the tested range. More
importantly, replacing increasingly large fractions of the candidate set with
same-cluster distractors produced only a moderate population-level decline, with
retrieval remaining well above chance even when the search space was dominated by
semantically related alternatives. Thus, the representation cannot be explained
solely by coarse category information; it retains information that distinguishes
individual stimuli within semantic neighbourhoods. Consistent with this result,
nearest-neighbour structure in brain space overlapped with that of CLIP more than
expected by random correspondence across neighbourhood scales. The absolute overlap was modest, however, indicating partial preservation rather than a wholesale reproduction of CLIP's local topology. Together, these results support partial preservation of meaningful visual-semantic neighbourhood structure from CLIP, while the incomplete overlap leaves room for prediction error, regularization, measurement noise, and potentially distinct representational structure.

The neuroanatomical analyses further ground this shared geometry in recognizable
cortical organization. Early and ventral visual systems formed the most consistent
backbone across datasets and participants, whereas motion and parietal-attention
systems contributed more selectively across stimulus categories. Cluster-specific
response maps, saliency and ablation therefore point to a representation that is
both shared and compositional; a common set of visual systems supports decoding
broadly, while the relative contribution of additional networks changes with visual
content. The convergence between saliency and ablation is particularly useful
because the two analyses probe different aspects of the model. Gradient-based
saliency identifies inputs to which an individual prediction is locally sensitive,
whereas ablation measures the loss of retrieval performance when an entire network
is removed. Their agreement is consistent with category-associated input dependence in the fitted mapping, but the two analyses share the same model and data and do not independently establish biological category selectivity.

Several limitations define the scope of these conclusions. First, although participant-agnostic models support target-dataset holdout retrieval, we have not yet quantified how efficiently limited participant-specific data could improve upon this initial representation. Comparing zero-shot transfer with few-shot estimation of a new participant embedding while freezing the shared encoder would clarify how much
individual adaptation remains beneficial. Second, CLIP provides the supervisory
target and consequently constrains the geometry that the model can recover. Although our analyses show that CLIP-like semantic relationships survive prediction from fMRI, they do not establish that this geometry is uniquely or completely representative of human visual organization. Comparisons with behaviour-derived similarity spaces, alternative self-supervised vision models and other multimodal representations could separate structure imposed by the target model from structure consistently supported by human neural data. A stronger test would compare raw-response, pre-projection and output geometries on held-out stimuli against CLIP, independently pretrained visual models and human similarity judgments, accounting for measurement reliability. THINGS behavioural judgments and their behaviour-only similarity embedding provide one such reference \cite{hebart2023things}; CLIP-based predictions that extend this embedding to new images \cite{contier2024distributed} should be distinguished from direct behavioural measurements. Alternative cluster definitions would test sensitivity to the grouping scheme, whereas alternative supervision would require retraining the decoder. Third, the source datasets remain highly imbalanced in
participants, trials and stimulus composition, making it difficult to isolate
diversity from scale. Finally, network masking and gradient attribution characterize the fitted decoder and can reflect group size, redundant inputs, and distribution shift. Biological interpretation would additionally require attribution validation and perturbation controls \cite{adebayo2018sanity}; these analyses do not establish causal necessity of the corresponding brain systems.

More broadly, this work suggests a different strategy for scaling computational
models of human brain activity. Progress need not depend exclusively on collecting a
single, uniformly acquired dataset of ever-increasing size. Independently acquired
experiments contain shared information that can become computationally compatible
when anatomical correspondence, cross-participant structure and stimulus semantics are
modelled jointly. Our target-dataset holdout results support decoding without fitting participant embeddings to the target data. Extending this framework to few-shot participant adaptation,
behaviourally grounded targets and other sensory or cognitive domains could provide a
route towards neural models that improve as heterogeneous datasets accumulate---
transforming cross-study variation from an obstacle to scaling into a source of
transferable structure.

\section*{Methods}
\label{sec:methods}

\subsection*{MOSAIC dataset and harmonization}
\label{sec:mosaic_data}

\paragraph{Dataset aggregation.}
We used MOSAIC, a harmonized collection of eight independently acquired
visual-fMRI datasets \cite{lahner2025mosaic}: the Natural Scenes Dataset
(NSD), THINGS-fMRI (THINGS), Deeprecon, BOLD5000, the BOLD Moments Dataset
(BMD), Generic Object Decoding (GOD), the Human Actions Dataset (HAD), and
the Natural Object Dataset (NOD). Together, these datasets span substantial
variation in acquisition protocol, scanner, stimulus domain, experimental
design, and amount of data collected per participant. After harmonization and
cross-dataset filtering, MOSAIC contains 93 dataset-specific participant entries,
430,007 single-trial fMRI--stimulus pairs, and 162,839 unique stimuli.

The curated splits comprise 144,360 naturalistic training stimuli associated
with 353,924 single-trial responses, 18,145 naturalistic test stimuli
associated with 68,011 responses, and 334 artificial test stimuli associated
with 8,072 responses. The experiments reported here use the naturalistic
training and validation partitions. HAD contributes to model training but is not included in the retrieval evaluation.

A dataset-specific participant entry is treated as a distinct modelling unit.
Accordingly, participant identity is defined jointly by dataset and participant label.
For example, \texttt{sub-01} from two different source datasets is assigned
two different participant indices. This convention allows the model to accommodate
dataset-specific differences in acquisition and recording conditions even when
participant labels coincide.

\paragraph{Cross-dataset split construction and leakage control.}
The scale of MOSAIC creates an important potential source of leakage; the same
or highly similar visual stimulus can occur in more than one constituent
dataset. MOSAIC therefore constructs its splits at the level of stimulus
identity rather than individual fMRI trials. Identical stimuli assigned to
conflicting source-dataset splits are first reconciled so that all observations
of a stimulus remain in the same partition.

A second filtering stage removes near-duplicate visual content. Training
stimuli are compared with test stimuli using cosine similarity between DreamSim
embeddings \cite{fu2023dreamsim}; training stimuli exceeding a similarity
threshold of $0.8196$ to a test stimulus are excluded. This procedure removed
1,186 training stimuli. Consequently, repeated observations of the same
stimulus---including observations obtained from different participants or
datasets---remain together, while identical or highly similar images cannot
appear on opposite sides of the train--test boundary.

\paragraph{fMRI preprocessing and single-trial response estimation.}
Rather than combining the heterogeneous preprocessing products released with the original datasets, MOSAIC reprocesses the constituent studies through a common pipeline \cite{lahner2025mosaic}. Functional data were preprocessed using fMRIPrep v23.2.0 \cite{esteban2019fmriprep} and registered to the HCP fsLR32k cortical surface space. Single-trial response amplitudes were estimated using GLMsingle \cite{prince2022glmsingle}, with dataset-specific treatment of
acquisition timing, stimulus duration, repeated presentations, and nuisance regressors.

Within each session or session group, beta estimates were normalized using statistics estimated exclusively from training trials. The resulting training-derived mean and standard deviation were then applied unchanged to both training and validation responses. No spatial smoothing was applied.
This harmonization step places measurements from otherwise heterogeneous
experiments into a common anatomical and statistical representation while
preventing information from the validation partition from entering the
normalization procedure.

\subsection*{ROI representation and model input}
\label{sec:roi_representation}

Each single-trial response was represented using $N=379$ anatomically defined regions of interest (ROIs): 360 cortical parcels from the HCP-MMP1.0 parcellation \cite{glasser2016multi}, comprising 180 parcels per hemisphere, together with 19 subcortical regions. This ROI-based representation provides an anatomically corresponding sequence across datasets while retaining the fine-grained activation pattern within each region.

For trial $t$ from dataset-specific participant $i$, the response in ROI $r$ is

\begin{equation}
    R_{t,r}^{(i)}
    =
    \left[
        \beta_{t,r,1}^{(i)},\ldots,
        \beta_{t,r,m_r}^{(i)}
    \right]
    \in \mathbb{R}^{m_r},
    \label{eq:roi_response}
\end{equation}

where $\beta_{t,r,k}^{(i)}$ denotes the single-trial beta coefficient of
cortical vertex or subcortical voxel $k$ within ROI $r$. Because anatomical
regions contain different numbers of vertices or voxels, $m_r$ varies across
ROIs. The model therefore does not assume that every ROI has the same raw
input dimensionality.

Before tokenization, an additional normalization was performed independently
within each ROI and trial. Non-finite values were first replaced with zero.
The resulting vector was then standardized across its constituent vertices or
voxels and clipped to $[-5,5]$:

\begin{equation}
    \widehat{R}_{t,r}^{(i)}
    =
    \operatorname{clip}
    \left(
        \frac{
            R_{t,r}^{(i)}
            -
            \mu\!\left(R_{t,r}^{(i)}\right)
        }{
            \sigma\!\left(R_{t,r}^{(i)}\right)+10^{-6}
        },
        -5,5
    \right).
    \label{eq:roi_normalization}
\end{equation}

This operation standardizes the distribution within each ROI while preserving
the relative spatial pattern of activation across its vertices or voxels.
Clipping reduces the influence of extreme coefficients. Spatial centering and scaling remove ROI-wide mean and scale information from each trial, so the decoder primarily uses within-ROI spatial patterns rather than absolute regional response amplitudes. Importantly, this
per-trial, within-ROI normalization is distinct from the training-derived
normalization applied during MOSAIC preprocessing.

\subsection*{Shared brain encoder with optional participant conditioning}
\label{sec:model}

The central modelling objective was to separate variation that is specific to
an ROI or participant from the parameters used to learn structure shared across the
entire collection. We therefore first map each anatomically corresponding ROI
into a common token space and only then apply a Transformer whose parameters
are shared across all 93 participant entries and all eight datasets
(Fig.~\ref{fig:schematic}).

\paragraph{ROI tokenization.}
Because the raw dimensionality $m_r$ differs across ROIs, each region has its
own two-layer multilayer perceptron (MLP) tokenizer. For simplicity, we omit
the trial index below. The standardized response from ROI $r$ of participant $i$
is mapped to

\begin{equation}
    E_r^{(i)}
    =
    f_r\!\left(\widehat{R}_r^{(i)}\right)
    =
    W_{r,2}\,
    \operatorname{GELU}
    \left(
        W_{r,1}\widehat{R}_r^{(i)}+b_{r,1}
    \right)
    + b_{r,2},
    \label{eq:roi_tokenizer}
\end{equation}

where
$W_{r,1}\in\mathbb{R}^{d\times m_r}$,
$W_{r,2}\in\mathbb{R}^{d\times d}$, and the common token dimension is
$d=512$. Thus, the ROI-specific tokenizers absorb differences in the number of
measurements available within each anatomical region while presenting the
shared encoder with a fixed-dimensional sequence.

\paragraph{ROI identity and participant conditioning.}
Each token is augmented with two learned embeddings: an ROI identity embedding
$P_r\in\mathbb{R}^{d}$ and a participant embedding
$S_i\in\mathbb{R}^{d}$,

\begin{equation}
    \widetilde{E}_r^{(i)}
    =
    E_r^{(i)} + P_r + S_i,
    \qquad r=1,\ldots,379.
    \label{eq:conditioned_roi_tokens}
\end{equation}

The ROI embedding allows anatomically distinct tokens to retain their identity after entering the shared Transformer. The same participant embedding $S_i$ is broadcast to all 379 ROI tokens from participant $i$, providing a low-dimensional mechanism through which the network can account for participant- and dataset-specific variation without introducing a separate encoder for each individual. Because participant identities are defined jointly by dataset and participant label, identically named participants from different datasets receive different embeddings.

Participant conditioning was included to provide a lightweight mechanism for modelling participant-specific variation that can remain after anatomical harmonization. Related cross-participant decoding approaches similarly use participant-specific functional alignment or adaptation before applying shared decoding components \cite{ferrante2024through,scotti2024mindeye2}. In our
model, this role is restricted to the additive embedding $S_i$; the ROI
tokenizers, Transformer encoder, attention pooling and projection head remain shared.

We therefore evaluate two operating modes. In the \emph{participant-conditioned}
mode used for the standard retrieval and interpretability analyses, the learned
embedding $S_i$ is added to every ROI token for participant $i$. In the
\emph{participant-agnostic} mode used for zero-shot transfer, participant embeddings
are disabled, so that
\begin{equation}
    \widetilde{E}_r
    =
    E_r + P_r,
\end{equation}
and no participant-specific embedding is required. This second mode permits
direct evaluation on participants whose identities were never represented
during training.

\paragraph{Shared Transformer encoder.}
The conditioned sequence

\begin{equation}
    \widetilde{\mathbf{E}}^{(i)}
    =
    \left[
        \widetilde{E}_1^{(i)},\ldots,
        \widetilde{E}_{379}^{(i)}
    \right]
\end{equation}

is processed by a four-layer pre-normalization Transformer encoder
\cite{vaswani2017attention}. Each layer contains four self-attention heads,
a token dimension of 512, a feed-forward dimension of 2,048, GELU
activations, and dropout of $0.1$. The Transformer parameters are shared
across every ROI sequence, participant, and dataset. A final layer normalization
is applied to the resulting contextualized ROI representations,
$H_r^{(i)}\in\mathbb{R}^{512}$.

This design allows the participant embedding to modulate the representation
presented to the shared network while requiring the Transformer itself to
learn interactions among anatomically corresponding regions that are useful
across the heterogeneous training collection.

\subsection*{Attention pooling and alignment with CLIP}
\label{sec:clip_alignment}

\paragraph{Attention pooling.}
The Transformer produces one contextualized representation for each ROI. We
combine these representations using learned attention pooling rather than
assuming that every region contributes equally to every trial. A two-layer
scoring network assigns

\begin{align}
    a_r^{(i)}
    &=
    w_2^\top
    \operatorname{GELU}
    \left(
        W_1\operatorname{LN}(H_r^{(i)})+b_1
    \right)
    +b_2,\\
    \alpha_r^{(i)}
    &=
    \frac{\exp(a_r^{(i)})}
    {\sum_{q=1}^{379}\exp(a_q^{(i)})},
\end{align}

where the scoring network has a hidden dimension of 256. The trial-level brain
representation is then

\begin{equation}
    B^{(i)}
    =
    \sum_{r=1}^{379}
    \alpha_r^{(i)}H_r^{(i)}.
    \label{eq:attention_pool}
\end{equation}

The pooled representation is passed through a projection head consisting of
layer normalization, a $512\rightarrow512$ linear transformation, GELU, and
a second $512\rightarrow512$ linear transformation. The resulting vector,

\begin{equation}
    \widehat{C}^{(i)}\in\mathbb{R}^{512},
\end{equation}

is the model prediction in visual-semantic embedding space.

\paragraph{CLIP target representations.}
Target stimulus representations were precomputed using the image encoder of
OpenAI CLIP ViT-B/32 \cite{radford2021learning}. For static-image stimuli,
the presented image was passed directly through the frozen CLIP image encoder,
yielding a 512-dimensional target embedding
$C^{(j)}\in\mathbb{R}^{512}$.

For datasets containing dynamic video stimuli, including BMD and HAD, each
video was represented by its temporally central frame. Specifically, the frame
at, or nearest to, the temporal midpoint of the presented video clip was
extracted and passed through the same CLIP ViT-B/32 image encoder. The
resulting 512-dimensional embedding was used as the visual-semantic target for
all fMRI responses associated with that video stimulus. This provided a single
fixed-dimensional image-space representation for each dynamic stimulus while
keeping the target representation consistent with that used for the
static-image datasets. Accordingly, this target evaluates correspondence to a representative frame and does not separately establish decoding of temporal dynamics or actions.

The CLIP encoder remained frozen throughout training; gradients were propagated
only through the brain encoder, attention-pooling module, and projection head.

\subsection*{Multi-positive contrastive learning}
\label{sec:contrastive_training}

When repeated stimuli are sampled together, several fMRI observations within a training batch can correspond to the same stimulus, whether within or across participants and datasets. These repetitions may arise
within one participant, across participants, or across datasets. Treating all other
samples in the batch as conventional contrastive negatives would therefore
incorrectly push valid representations of the same stimulus apart.

We instead optimized symmetric cross-entropy with a uniform target over all same-stimulus observations in the distributed batch. For brain
sample $i$, let $\mathcal{P}(i)$ denote all samples in the distributed batch
with the same stimulus identity. The target distribution is

\begin{equation}
    y_{ij}
    =
    \begin{cases}
        1/|\mathcal{P}(i)|, & j\in\mathcal{P}(i),\\
        0,                  & \text{otherwise}.
    \end{cases}
    \label{eq:multipositive_target}
\end{equation}

Predicted brain embeddings and target CLIP embeddings are first
$\ell_2$-normalized. Brain-to-CLIP similarity logits are then

\begin{equation}
    \ell^{\mathrm{B2C}}_{ij}
    =
    \gamma\,
    \widehat{C}^{(i)\top}C^{(j)},
    \label{eq:b2c_logits}
\end{equation}

where $\gamma$ is a learned logit scale initialized to $1/0.07$ and capped at
15. The brain-to-CLIP objective is

\begin{equation}
    \mathcal{L}_{\mathrm{B2C}}
    =
    -\frac{1}{B}
    \sum_{i=1}^{B}
    \sum_j
    y_{ij}
    \log
    \frac{
        \exp(\ell^{\mathrm{B2C}}_{ij})
    }{
        \sum_k\exp(\ell^{\mathrm{B2C}}_{ik})
    }.
    \label{eq:b2c_loss}
\end{equation}

We analogously compute
$\mathcal{L}_{\mathrm{C2B}}$ by reversing the query and target roles and use
the symmetric objective

\begin{equation}
    \mathcal{L}_{\mathrm{contrastive}}
    =
    \frac{1}{2}
    \left(
        \mathcal{L}_{\mathrm{B2C}}
        +
        \mathcal{L}_{\mathrm{C2B}}
    \right).
    \label{eq:contrastive_loss}
\end{equation}

Embeddings and stimulus identifiers are gathered across GPUs before the loss is
computed, so positive and negative relationships are defined across the full
distributed batch rather than independently on each device. The contrastive
loss was assigned a weight of $0.5$. The general implementation also supports
masked autoencoding; this branch was disabled for every experiment reported
here.

\subsection*{Model optimization}
\label{sec:optimization}
Models were optimized with AdamW using an initial learning rate of $10^{-4}$ and weight decay of $10^{-3}$. Bias parameters, normalization parameters, and the learned contrastive logit scale were excluded from weight decay. The learning rate followed a per-epoch cosine-annealing schedule. Training used 256 samples per GPU across four GPUs with distributed data parallelism, giving an effective batch size of 1{,}024. Embeddings were gathered across GPUs so that each contrastive loss term was computed against all 1{,}024 samples in the global batch. Models were trained in 32-bit precision with gradient-norm clipping at 1.0 and random seed 42.

Early stopping and checkpoint selection were based on the held-out evaluation split. The monitored quantity was a minibatch-level proxy; Top-5 retrieval within each per-GPU minibatch of 256 samples, averaged over the epoch. This proxy is distinct from the full-gallery retrieval metrics reported in the Results. Training stopped when this proxy failed to improve by at least $10^{-4}$ for five consecutive epochs.
Additional checkpoint analysis can be found in section \ref{checkpoint}.

\subsection*{Stimulus retrieval}
\label{sec:retrieval_evaluation}

We evaluated whether the predicted brain representation identified the
stimulus associated with a held-out fMRI response. Predicted brain embeddings and candidate CLIP embeddings were $\ell_2$-normalized, making their inner product equivalent to cosine similarity. For each held-out brain query $q$, candidate stimuli were ranked by decreasing brain--CLIP similarity.

Repeated candidate entries were collapsed by stimulus identity before ranking; the accuracy below averages success over the $Q$ evaluated fMRI queries. Retrieval therefore asks which \emph{unique stimulus} is represented by a given brain response rather than rewarding repeated entries of the same target. Unique-stimulus Top-$K$ accuracy was defined as

\begin{equation}
    \operatorname{Top}\text{-}K
    =
    \frac{1}{Q}
    \sum_{q=1}^{Q}
    \mathbf{1}
    \left[
        \operatorname{rank}_q(y_q)\leq K
    \right],
    \label{eq:topk}
\end{equation}

where $y_q$ is the identity of the true stimulus and $Q$ is the number of
evaluated fMRI queries. Unless otherwise stated, we report Top-10 accuracy.
For a random ranking over $N_c$ unique candidates, chance Top-10 performance is

\begin{equation}
    p_{\mathrm{chance}}=\min\!\left(\frac{10}{N_c},1\right).
\end{equation}

We considered two candidate spaces. In the \emph{global} evaluation, each
query was ranked against all unique validation stimuli pooled across the seven evaluation datasets. This is the most demanding setting because candidates can originate from entirely different experiments and stimulus distributions. In the \emph{dataset-restricted} evaluation, candidate stimuli were limited to the dataset from which the query originated. The latter provides a complementary measure that removes cross-dataset distractors and has a dataset-dependent
chance level determined by the number of unique candidate stimuli.

\subsection*{Participant-matched baseline comparison}
\label{sec:baseline_methods}

Following Benchetrit et al.~\cite{benchetrit2026boosting}, our MindEye-style baseline predicts a single pooled CLIP embedding for retrieval, without the diffusion prior or image reconstruction components of the full MindEye framework \cite{scotti2023mindeye}.
We compared \model with ridge regression and this MindEye-style
pooled-CLIP decoder on the subset of participant entries for which matched results were available for all three approaches; four from BOLD5000 and four from NSD. All comparisons use unique-stimulus Top-10 retrieval accuracy. Dataset-level and overall summaries are reported as the mean $\pm$ standard deviation across matched participants.




\subsection*{Semantic organization of the stimulus space}
\label{sec:semantic_clustering}

To examine whether retrieval depended systematically on stimulus content, we
partitioned the unique validation stimuli according to their CLIP
representations. Target embeddings were $\ell_2$-normalized and clustered
using $k$-means with $k=40$. A two-dimensional t-SNE projection was used only
for visualization and played no role in defining clusters or computing the
subsequent statistics. Semantic category names are descriptive annotations of the embedding-defined clusters.

For each semantic cluster, retrieval accuracy was obtained by averaging
query-level Top-10 success over queries whose target belonged to that cluster.
To avoid selecting categories after inspecting their identities, clusters were
first ordered solely by retrieval accuracy and every fourth cluster in this
ordering was retained. The selected ranks were

\begin{equation}
    1,5,9,\ldots,37,
\end{equation}

yielding ten semantic clusters spanning the observed range from high to low
retrieval performance. These same ten clusters were used for subsequent
representational-geometry, ablation, and saliency analyses.

\subsection*{Functional-network ablation}
\label{sec:network_ablation_methods}

To quantify the effect of masking large-scale anatomical and functional input groups on model retrieval, the 379 ROIs were assigned to nine functional or anatomical groups: early visual, motion visual, ventral visual/object, parietal attention, frontal/control, temporal/auditory, cerebellar, subcortical, and a residual ``Other'' group. 
Each group was ablated independently at inference by removing its tokens from the model's computation. Ablated tokens were excluded as keys in every self-attention layer of the encoder and received zero weight in the
attention-pooling readout, so pooling weights were renormalized over the
remaining tokens. Because the encoder is pre-normalized and acts on each
token independently except through attention, and because ROI identity
embeddings are the only positional signal, this is equivalent to deleting the ablated tokens from the input set. It differs from zeroing the input signal; a zero-valued input would still produce a token through the tokenizer biases and the added ROI and subject embeddings. Model parameters, subject embeddings of the retained tokens, and the retrieval candidate pool were unchanged.

For participant $i$ and network $g$, we defined the retrieval drop as

\begin{equation}
    D_{i,g}
    =
    A_i^{\mathrm{baseline}}
    -
    A_{i,g}^{\mathrm{ablated}},
    \label{eq:ablation_drop}
\end{equation}

where accuracy is measured in percentage points. Positive values therefore
indicate impaired retrieval following ablation. Participant identities were
defined using the joint identifier \texttt{dataset::participant}; if multiple
entries existed for the same participant--network combination, their retrieval
drops were averaged before inference.

For each network, we used a one-sided Wilcoxon signed-rank test of a positive shift in retrieval drops. Interpreting the following hypothesis as a median shift requires an approximately symmetric distribution of independent differences:

\begin{equation}
    H_0:
    \operatorname{median}(D_g)\leq0,
    \qquad
    H_1:
    \operatorname{median}(D_g)>0.
\end{equation}

Zero-valued differences were omitted according to the Wilcoxon
\texttt{zero\_method=wilcox} convention. The resulting $P$ values were
corrected across the nine network tests using the Benjamini--Hochberg
false-discovery-rate procedure with $\alpha=0.05$.

We estimated 95\% percentile-bootstrap confidence intervals for both the mean
and median retrieval drop using 20,000 resamples with random seed 42. As an
effect-size measure, we additionally computed the signed-rank rank-biserial
correlation,

\begin{equation}
    r_{\mathrm{rb}}
    =
    \frac{W^{+}-W^{-}}
         {W^{+}+W^{-}},
    \label{eq:rank_biserial}
\end{equation}

where $W^{+}$ and $W^{-}$ are the sums of ranks associated with positive and
negative non-zero differences, respectively.

Differences among networks were additionally assessed using a Friedman
repeated-measures omnibus test over participants with complete observations for
all nine groups. We report Kendall's coefficient of concordance as the
corresponding omnibus effect size,

\begin{equation}
    W
    =
    \frac{\chi^2_{\mathrm{Friedman}}}
         {n(k-1)},
    \label{eq:kendall_w}
\end{equation}

where $n$ is the number of complete participants and $k=9$.

For visualization, ablation effects are plotted using the reverse sign,

\begin{equation}
    \Delta A_{i,g}
    =
    A_{i,g}^{\mathrm{ablated}}
    -
    A_i^{\mathrm{baseline}}
    =
    -D_{i,g},
\end{equation}

such that negative plotted values denote impaired retrieval. Pooled effects
weight individual fMRI queries equally, whereas participant-level summaries weight
participants equally; the corresponding unablated baseline accuracies can therefore
differ.

\subsection*{Zero-shot cross-dataset transfer}
\label{sec:zero_shot_methods}

For all zero-shot curriculum experiments, participant embeddings were disabled throughout model training and evaluation. For each target dataset, the dataset was excluded in its entirety, including all of its participants and stimuli.
Consequently, no target dataset's participant embeddings or other parameters were learned or estimated, and no training data from the target dataset were available to the model. The target evaluation set remained fixed while the source training pool was expanded cumulatively. At each curriculum step, one additional dataset was added to the training pool; the target dataset was never introduced during this sequence.

Let $A_t$ denote target-dataset Top-10 retrieval accuracy after curriculum step
$t$. To place datasets with different absolute retrieval accuracies on a common
scale, performance was expressed relative to the initial source-training
condition:

\begin{equation}
    \Delta_t^{(0)}
    =
    100
    \left(
        \frac{A_t}{A_0}-1
    \right).
    \label{eq:transfer_relative}
\end{equation}

The step-0 model therefore appears at $0\%$. Incremental annotations in the
figure report the change relative to the immediately preceding step,

\begin{equation}
    \Delta_t^{(\mathrm{prev})}
    =
    100
    \left(
        \frac{A_t}{A_{t-1}}-1
    \right),
    \label{eq:transfer_incremental}
\end{equation}

whereas the final annotation summarizes the total change between the first and
last curriculum conditions. This analysis therefore tests whether progressively broader cross-dataset
pretraining improves calibration-free transfer to entirely unseen participants
and datasets in the absence of participant-specific conditioning.

\subsection*{Semantic geometry of predicted brain representations}
\label{sec:semantic_geometry_methods}

Retrieval evaluates whether the true stimulus receives a sufficiently high rank, but does not reveal how the remaining candidates are organized around a brain prediction. We therefore tested whether the learned representation preserved graded semantic neighbourhood structure.

For every held-out brain query, candidate CLIP embeddings were partitioned into three classes: the matched stimulus; non-matching stimuli belonging to the same semantic cluster as the target; and stimuli belonging to different clusters. We computed the mean cosine similarity to each class and summarized the resulting ordering using two margins,

\begin{align}
    M_{\mathrm{identity}}
    &=
    s_{\mathrm{matched}}
    -
    s_{\mathrm{same}},
    \label{eq:identity_margin}\\
    M_{\mathrm{semantic}}
    &=
    s_{\mathrm{same}}
    -
    s_{\mathrm{different}}.
    \label{eq:semantic_margin}
\end{align}

The identity margin measures stimulus-specific information beyond semantic category membership, whereas the semantic margin measures preservation of broader category structure. Margins were summarized at both the semantic cluster and dataset levels. Cluster-level comparisons were corrected for multiple testing; hollow markers in the corresponding figure denote comparisons with FDR-adjusted $q\geq0.05$.

\subsection*{ROI saliency and semantic selectivity}
\label{sec:saliency_methods}

Finally, we asked whether different portions of the learned semantic space relied on different input regions. For each brain query $q$, we first computed the cosine similarity $s_q$ between the $\ell_2$-normalized predicted embedding and its matched CLIP embedding. ROI importance for region $r$ was then quantified by summing, across
its voxels, the absolute product of each input value $x_v$ and the
sensitivity of $s_q$ to that input, $\partial s_q/\partial x_v$. Attributions were computed with respect to the model input, for which the beta values within each ROI were z-scored across that ROI's voxels or vertices for every trial and clipped at
$\pm 5$. They therefore reflect the spatial pattern of activity within each ROI rather than its mean amplitude. The ROI-level attribution was defined as \[
S_{q,r} = \sum_{v\in r} \left| x_v \frac{\partial s_q}{\partial x_v}
\right|, \]
and each query's saliency vector was normalized to sum to one across ROIs, so that $S_{q,r}$ represents the fraction of total attribution assigned to ROI $r$.


For ROI $r$ and semantic cluster $c$, we quantified cluster selectivity by comparing the normalized ROI attribution $S_{q,r}$ for queries belonging to cluster $c$ with that for all queries outside the cluster:

\begin{equation}
    Z_{r,c}
    =
    \frac{
        \mu_{r,c}^{\mathrm{in}}
        -
        \mu_{r,c}^{\mathrm{out}}
    }{
        \sigma_{r,c}^{\mathrm{pooled}}
    },
    \label{eq:roi_selectivity}
\end{equation}

where $\mu_{r,c}^{\mathrm{in}}$ and $\mu_{r,c}^{\mathrm{out}}$ are the mean values of $S_{q,r}$ for queries inside and outside cluster $c$, respectively, and $\sigma_{r,c}^{\mathrm{pooled}}$ is their pooled within-group standard deviation. Thus, $Z_{r,c}$ is a standardized effect size (Cohen's $d$) where positive values indicate greater relative attribution to ROI $r$ for stimuli
in cluster $c$. It is used here as a descriptive selectivity measure rather than as a standard-normal test statistic. ROIs were ranked by $Z_{r,c}$, and the top $n_c = 20$ ROIs with positive selectivity were retained for each cluster.

Because the functional groups contain unequal numbers of ROIs, raw counts among these selected ROIs would favour larger groups. We therefore compared the observed count with that expected if selected ROIs were distributed in proportion to each group's representation in the atlas:

\begin{equation}
    \operatorname{Expected}_{g,c}
    =
    n_c\frac{N_g}{N},
    \label{eq:expected_network_count}
\end{equation}

where $N_g$ is the total number of eligible ROIs assigned to functional group $g$, and $N = 379$ is the total number of eligible ROIs. We then defined the network-enrichment score as

\begin{equation}
    \operatorname{Enrichment}_{g,c}
    =
    \operatorname{Observed}_{g,c}
    -
    \operatorname{Expected}_{g,c},
    \label{eq:network_enrichment}
\end{equation}

where $\operatorname{Observed}_{g,c}$ is the number of selected ROIs from group $g$. Positive values therefore indicate that a functional group contributes more cluster-selective ROIs than expected from its overall representation in the atlas, whereas negative values indicate fewer than expected.

\subsection*{Cross-dataset semantic geometry and retrieval robustness}
\label{sec:generalization_methods}

\paragraph{Hierarchical geometry within the learned brain space.}
To test whether stimulus identity and semantic structure were preserved directly
within the learned brain representations, we compared predicted brain embeddings
across participant entries. Because the same stimulus can have multiple fMRI
observations within a participant, embeddings were first averaged for each
participant--stimulus combination and $\ell_2$-normalized. We then considered stimuli observed by at least two distinct participant entries and formed pairs of embeddings corresponding to the same stimulus. To prevent stimuli seen by many participants from disproportionately influencing the analysis, at most ten participant pairs were sampled per stimulus.

For each same-stimulus participant pair $(i,j)$, we computed the cosine similarity
between their normalized brain embeddings. We constructed two control conditions
using stimuli available for the corresponding participants. For the same-cluster
condition, the embedding from participant $i$ was compared with up to five randomly
sampled embeddings from participant $j$ belonging to the same semantic cluster but
corresponding to different stimuli; the reverse comparison, from participant $j$ to
participant $i$, was computed analogously, and the two directional means were averaged.
The different-cluster condition was constructed identically, except that control
stimuli were sampled from semantic clusters different from that of the target.
Sampling was without replacement.

For each participant pair, similarities were then averaged across all shared stimuli,
yielding three estimates:
$s_{\mathrm{same\ stimulus}}$,
$s_{\mathrm{same\ cluster}}$, and
$s_{\mathrm{different\ cluster}}$. We quantified hierarchical organization using

\begin{align}
    M_{\mathrm{identity}}
    &=
    s_{\mathrm{same\ stimulus}}
    -
    s_{\mathrm{same\ cluster}},\\
    M_{\mathrm{semantic}}
    &=
    s_{\mathrm{same\ cluster}}
    -
    s_{\mathrm{different\ cluster}}.
\end{align}

Positive $M_{\mathrm{identity}}$ indicates convergence across brains that is specific to stimulus identity beyond semantic category, whereas positive $M_{\mathrm{semantic}}$ indicates preservation of broader semantic structure.
Both margins were tested across participant pairs using one-sided Wilcoxon signed-rank tests with the alternative hypothesis that the margin was greater than zero.
For visualization, participant pairs were additionally classified according to whether they originated from the same or different datasets. Group means and 95\% confidence intervals were estimated by bootstrapping the participant-pair estimates with 5,000 resamples. These pair-level summaries do not account for dependence induced by participants or stimuli shared across pairs, and therefore do not establish population-level uncertainty over independent people and stimuli. Inference that generalizes over people and stimuli requires a design that respects both sampling factors \cite{schutt2023inference}.

\paragraph{Cross-dataset correspondence of semantic representational geometry.}
We tested whether the relative organization of semantic categories in the learned brain space was preserved across datasets. To avoid weighting stimuli according to the number of repeated fMRI observations, we first averaged all predicted brain
embeddings corresponding to the same unique stimulus within each dataset and $\ell_2$-normalized the resulting stimulus-level embedding. Within each dataset and semantic cluster, these stimulus-level embeddings were then averaged to obtain a dataset-specific cluster centroid and normalized again. Clusters containing fewer
than five unique stimuli were excluded.

For each pair of datasets, we retained semantic clusters represented in both datasets and required at least five shared clusters. A semantic representational dissimilarity matrix (RDM) was constructed separately for each dataset using cosine
distance between its cluster centroids,
\begin{equation}
    D_{ab}
    =
    1-\mathbf{c}_{a}^{\top}\mathbf{c}_{b},
\end{equation}
where $\mathbf{c}_{a}$ and $\mathbf{c}_{b}$ are $\ell_2$-normalized centroids for
semantic clusters $a$ and $b$. Cross-dataset correspondence was quantified as the
Spearman correlation, $\rho$, between the upper-triangular entries of the two RDMs.

Statistical significance was assessed using a Mantel-style permutation test with
2,000 permutations. At each permutation, the cluster ordering of one RDM was
randomly permuted identically across its rows and columns before recomputing
$\rho$. The one-sided permutation $P$ value was calculated as
\begin{equation}
    P
    =
    \frac{
        1+\sum_{m=1}^{2000}
        \mathbf{1}
        \left[
            \rho_m^{\mathrm{null}}
            \geq
            \rho_{\mathrm{obs}}
        \right]
    }{
        2001
    }.
\end{equation}
Permutation $P$ values were corrected across all evaluable dataset pairs using
the Benjamini--Hochberg false-discovery-rate procedure. Dataset pairs with fewer
than five overlapping semantic clusters were not evaluated.

\paragraph{Candidate-pool scaling.}
To quantify how retrieval performance changes as the search space expands, we
evaluated candidate pools of 25, 50, 100, 250, 500, 1,000, 2,500, 5,000 and
10,000 unique stimuli, together with the full candidate pool when not already
included among these sizes. Rather than repeatedly subsampling candidate sets,
we computed the exact expected Top-10 accuracy for each brain query from its
rank in the full retrieval pool using the hypergeometric distribution.

For a query whose true stimulus has rank $r$ in a full pool containing
$N_{\mathrm{full}}$ candidates, there are $r-1$ distractors ranked above the
target among the $N_{\mathrm{full}}-1$ available distractors. For a reduced
candidate pool of size $N$, the target is accompanied by $N-1$ distractors
sampled uniformly without replacement. The probability that the target appears
within the Top-10 is therefore

\begin{equation}
    p_{\mathrm{Top10}}(N)
    =
    \Pr
    \left[
        X \leq 9
    \right],
    \qquad
    X
    \sim
    \operatorname{Hypergeom}
    \left(
        N_{\mathrm{full}}-1,\,
        r-1,\,
        N-1
    \right),
    \label{eq:candidate_pool_hypergeom}
\end{equation}

where $X$ is the number of sampled distractors that rank above the true
stimulus. Thus, the analysis gives the exact expected retrieval performance
under uniform candidate subsampling without requiring repeated random draws.

Query-level probabilities were first averaged within each participant at each
candidate-pool size. Dataset-specific curves show the mean of these
participant-level expectations within each dataset, whereas the overall curve
shows the mean across participants with a 95\% bootstrap confidence interval.
Random Top-10 performance for a candidate pool of size $N$ was

\begin{equation}
    p_{\mathrm{chance}}(N)
    =
    \min\left(\frac{10}{N},1\right).
    \label{eq:candidate_pool_chance}
\end{equation}

This analysis isolates the effect of increasing search-space size while
preserving the ranking structure observed in the full retrieval evaluation.

\paragraph{Semantic hard-negative retrieval.}
To test whether successful retrieval depended primarily on coarse semantic
category information, we constructed fixed-size candidate pools containing
progressively larger fractions of distractors from the same semantic cluster as
the target. Hard-negative fractions were $0$, $0.25$, $0.50$, $0.75$, and
$1.00$. The candidate-pool size was chosen as

\begin{equation}
N_{\mathrm{hard}}
=
\min\left(
200,
N_{\mathrm{min\ cluster}}
\right),
\end{equation}

where $N_{\mathrm{min\ cluster}}$ is the number of candidate stimuli in the
smallest semantic cluster. This ensured that even the $100\%$ hard-negative
condition could be formed without replacement for every query.

For each brain query, we used its similarity to the full candidate set to count
separately the number of same-cluster and different-cluster distractors that
ranked above the true stimulus. Let $N_s$ and $N_d$ denote the available
same-cluster and different-cluster distractors, and let $B_s$ and $B_d$ denote
the corresponding numbers that outrank the target. For a hard-negative fraction
$f$, the numbers of sampled same- and different-cluster distractors were

\begin{align}
n_s
&=
\operatorname{round}
\left[
(N_{\mathrm{hard}}-1)f
\right],\\
n_d
&=
(N_{\mathrm{hard}}-1)-n_s.
\end{align}

The numbers of sampled distractors that outrank the target therefore follow

\begin{equation}
X_s\sim
\operatorname{Hypergeom}(N_s,B_s,n_s),
\qquad
X_d\sim
\operatorname{Hypergeom}(N_d,B_d,n_d).
\end{equation}

Because the target is retrieved within the Top-10 whenever at most nine sampled
distractors outrank it, its exact success probability is

\begin{equation}
p_{\mathrm{Top10}}(f)
=
\sum_{x=0}^{9}
\Pr(X_s=x)
\Pr(X_d\leq 9-x).
\label{eq:hard_negative_top10}
\end{equation}

Thus, expected performance at each hard-negative fraction was obtained
analytically from the full retrieval ordering rather than by repeated candidate
resampling. Query-level probabilities were averaged within participant and then
summarized by dataset; the overall trajectory shows the mean across participants
with a 95\% bootstrap confidence interval. Because candidate-pool size remained
fixed across hard-negative conditions, random Top-10 chance was constant at

\begin{equation}
p_{\mathrm{chance}}
=
\min\left(
\frac{10}{N_{\mathrm{hard}}},
1
\right).
\end{equation}

This analysis progressively increases the semantic similarity of the distractor
set while holding its size fixed, providing a direct test of whether the learned
representation can discriminate individual stimuli within their broader
semantic neighbourhood.

\paragraph{Local neighbourhood preservation.}
To test whether the learned brain representation preserved the local topology of
the CLIP stimulus space, we compared $k$-nearest-neighbour sets around each unique
stimulus in the two embedding spaces. Because a stimulus can have multiple fMRI
observations, all predicted brain embeddings associated with the same stimulus
were first averaged and the resulting stimulus-level representation was
$\ell_2$-normalized. CLIP embeddings were aligned to the same unique-stimulus
ordering and likewise $\ell_2$-normalized.

For each stimulus, nearest neighbours were identified independently in brain and
CLIP space by cosine similarity, excluding the stimulus itself. We evaluated
neighbourhood sizes
$k\in\{5,10,20,50,100,250\}$. Local neighbourhood preservation was quantified as

\begin{equation}
O_k(i)
=
\frac{
\left|
\mathcal{N}^{\mathrm{brain}}_k(i)
\cap
\mathcal{N}^{\mathrm{CLIP}}_k(i)
\right|
}{k},
\label{eq:knn_overlap}
\end{equation}

where $\mathcal{N}^{\mathrm{brain}}_k(i)$ and
$\mathcal{N}^{\mathrm{CLIP}}_k(i)$ denote the $k$ nearest neighbours of stimulus
$i$ in the two spaces. Overlap fractions were averaged across all unique stimuli,
with 95\% bootstrap confidence intervals for the mean.

Under random correspondence between two neighbourhoods of size $k$ drawn from
the same $N-1$ possible non-self stimuli, the expected overlap fraction is

\begin{equation}
O_k^{\mathrm{random}}
=
\frac{k}{N-1},
\label{eq:knn_random_overlap}
\end{equation}

where $N$ is the total number of unique candidate stimuli. Observed overlap was
compared with this analytical random expectation across neighbourhood scales.
This analysis tests whether stimuli that are locally similar in CLIP space also
tend to occupy nearby positions in the learned brain-embedding space, beyond
what would be expected from random neighbourhood correspondence.

\section*{Data availability}

The source fMRI datasets and MOSAIC preprocessing resources are publicly available
through the repositories listed by the MOSAIC project \cite{lahner2025mosaic}.



\section*{Funding Statements}

This work was supported by Tether Operations, S.A. de C.V., which provided salary support for participating employees and the resources required to conduct the research. This work was also supported by the BRAINFEDERATION FISA-2022-00899, b the CROSSBRAIN project (European Union’s European Innovation Council under grant agreement No. 101070908; The BRAINSTORM PROJECT under Grant Agreement number 101099355).

\section*{Author contributions}

M.K. led the development of the study, including model design and implementation,
experimental design, model training and evaluation, quantitative analyses,
visualization, interpretation of results, and preparation of the manuscript.
M.T. contributed to methodology, model development, experimental analyses, and
interpretation of results. F.H. contributed substantially
to data analysis, visualization, interpretation of results, and manuscript writing
and revision. T.B. contributed to methodology and
manuscript revision. M.F. contributed to conceptualization, methodology,
interpretation of results, and manuscript revision. N.T. contributed to
conceptualization, interpretation of results, and manuscript revision.
All authors reviewed and approved the final manuscript.
\section*{Competing interests}
The authors declare no competing interests.





\clearpage
\setcounter{figure}{0}
\renewcommand{\thefigure}{S\arabic{figure}}

\renewcommand{\theHfigure}{S.\arabic{figure}}
\section*{Supplementary information}
\setcounter{table}{0}
\renewcommand{\thetable}{S\arabic{table}}
\renewcommand{\theHtable}{S.\arabic{table}}

\subsection*{Composition of the aggregated MOSAIC dataset}

Supplementary Table~\ref{tab:supp_datasets} summarizes the contribution of each
constituent dataset to MOSAIC before and after cross-dataset stimulus filtering.
The eight datasets together comprise 93 dataset-specific participant entries,
corresponding to 77 unique individuals. Because some individuals participated in
more than one source dataset, dataset-specific entries rather than unique people
define the modelling units used throughout this study. Filtering primarily removes
duplicate or highly similar stimuli across train--test partitions, reducing the
numbers of unique stimuli and associated trials while preserving the underlying
participant collection.

\begin{table}[h]
\centering
\caption{\textbf{Composition of the aggregated MOSAIC dataset before and after
cross-dataset stimulus filtering.}}
\label{tab:supp_datasets}
\small
\setlength{\tabcolsep}{2.5pt}
\begin{tabular}{lrrrrrr}
\toprule
Dataset & Entries & Naturalistic train & Naturalistic test & Artificial test & Total stimuli & Total trials \\
\midrule
BOLD5000 & 4 & 4,803 & 113 & 0 & 4,916 & 18,870 \\
BMD & 10 & 1,000 & 102 & 0 & 1,102 & 40,200 \\
GOD & 5 & 1,200 & 50 & 0 & 1,250 & 14,750 \\
Deeprecon & 3 & 1,200 & 50 & 50 & 1,300 & 24,360 \\
HAD & 30 & 16,200 & 5,400 & 0 & 21,600 & 21,600 \\
THINGS & 3 & 8,640 & 100 & 0 & 8,740 & 29,520 \\
NOD & 30 & 45,696 & 11,424 & 0 & 57,120 & 68,760 \\
NSD & 8 & 69,566 & 1,000 & 284 & 70,850 & 218,312 \\
\midrule
Agg. before filtering & 93 & 148,305 & 18,573 & 334 & 166,878 & 436,372 \\
Final after filtering & 93 & 144,360 & 18,145 & 334 & 162,839 & 430,007 \\
\bottomrule
\end{tabular}
\vspace{3pt}

\begin{minipage}{0.98\textwidth}
\footnotesize
\raggedright
The 93 dataset-specific participant entries correspond to 77 unique individuals;
16 individuals contributed recordings to more than one constituent dataset.
Filtering refers to cross-dataset stimulus reconciliation and near-duplicate
removal and therefore changes stimulus and trial counts, not the number of
participant entries.
\end{minipage}
\end{table}

\subsection*{Complete semantic organization of the validation stimulus space}

\begin{figure}[!h]
    \centering
    \includegraphics[width=\textwidth]{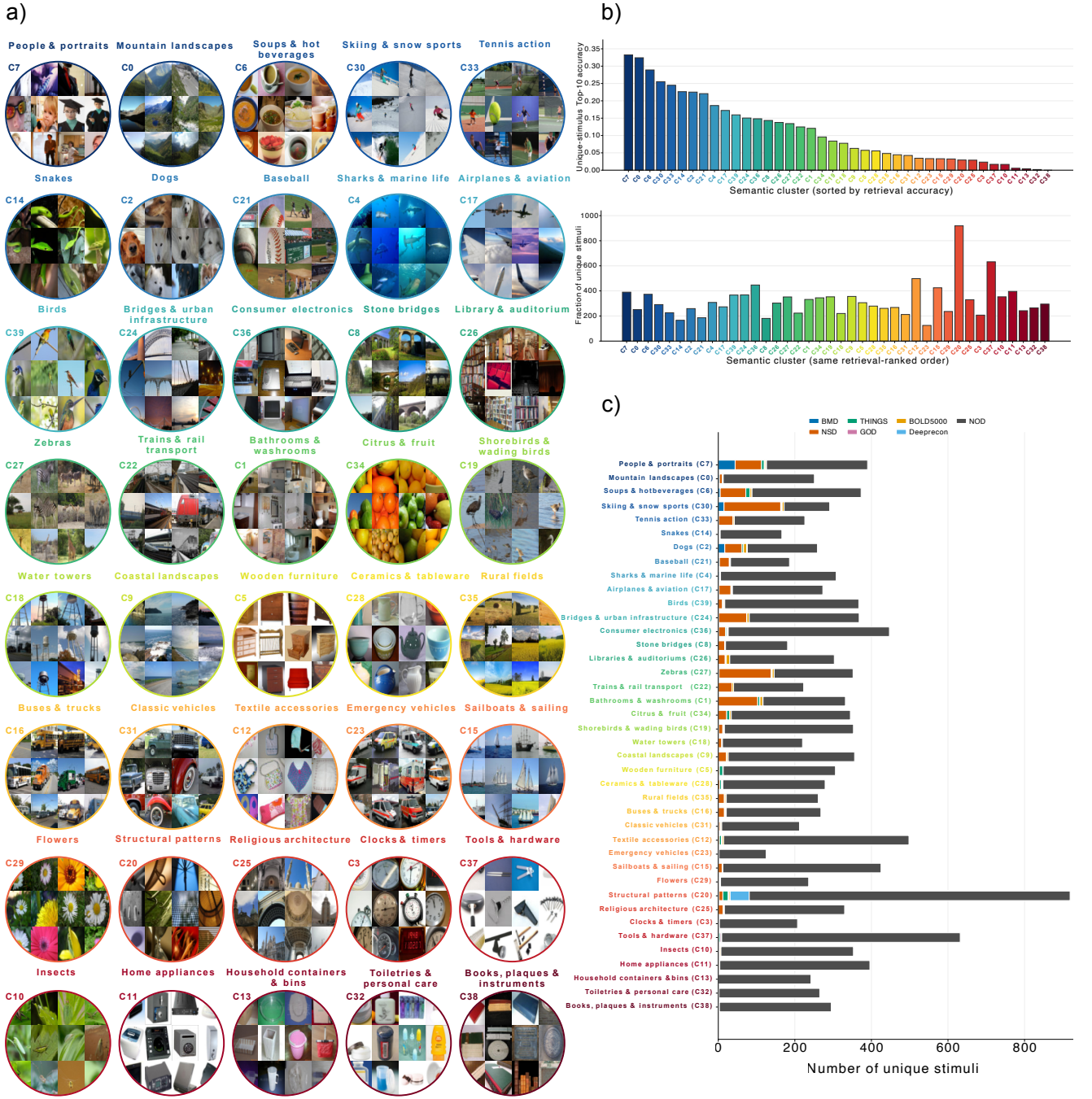}
    \caption{\captiontitle{Complete semantic organization and retrieval performance
    across all 40 stimulus clusters.}
    \textbf{a,} Representative stimuli from each of the 40 semantic clusters
    obtained by applying $k$-means to $\ell_2$-normalized CLIP embeddings of
    unique validation stimuli. Clusters are arranged from higher to lower
    Top-10 retrieval accuracy, with labels indicating the descriptive semantic
    category and arbitrary $k$-means cluster identifier (C0--C39).
    \textbf{b,} Unique-stimulus Top-10 retrieval accuracy (top) and number of
    unique stimuli (bottom) for all 40 clusters in the same retrieval-ranked
    order. The colour progression follows retrieval rank and is used
    consistently across panels.
    \textbf{c,} Dataset composition of each semantic cluster, shown in the same
    retrieval-ranked order as in \textbf{a,b}. Horizontal stacked bars indicate
    the number of unique stimuli contributed by each evaluation dataset.
    Cluster identifiers are arbitrary and are unrelated to retrieval rank.
    }
    \label{fig:supp_all_clusters}
\end{figure}

To complement the representative clusters shown in the main text, we examined
the complete set of 40 semantic clusters defined from normalized CLIP embeddings
of the unique validation stimuli (Supplementary
Fig.~\ref{fig:supp_all_clusters}). The clusters span a broad range of visual
content, including people, animals, actions, landscapes, vehicles, architecture
and manufactured objects (Supplementary Fig.~\ref{fig:supp_all_clusters}a).
Retrieval accuracy varied substantially across this full semantic spectrum and
did not simply follow the number of stimuli assigned to each cluster
(Supplementary Fig.~\ref{fig:supp_all_clusters}b). Cluster composition also
varied across source datasets, with many semantic categories represented by
stimuli from multiple datasets (Supplementary
Fig.~\ref{fig:supp_all_clusters}c). Together, these results show that the
content-dependent retrieval differences observed in the main analysis extend
across the complete stimulus taxonomy rather than being specific to the ten
clusters selected for detailed neuroanatomical analysis.

\subsection*{Semantic geometry and cortical contributions across all stimulus clusters}

\begin{figure}[!h]
    \centering
    \includegraphics[width=\textwidth]{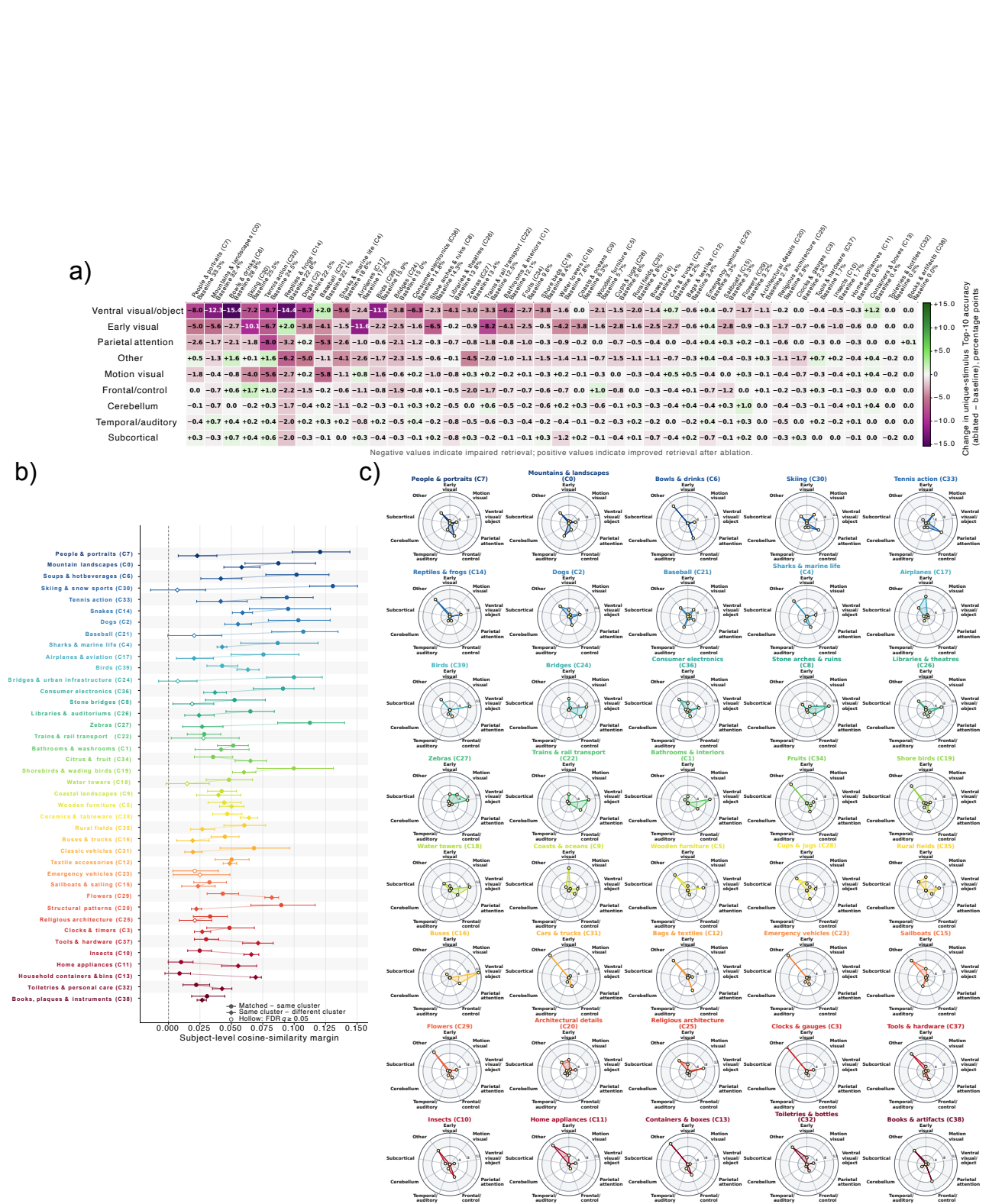}
    \caption{\captiontitle{Semantic geometry and cortical contributions across all
    40 stimulus clusters.}
    \textbf{a,} Cluster-specific effects of functional-network ablation on
    unique-stimulus Top-10 retrieval accuracy. Columns show all 40 semantic
    clusters in retrieval-ranked order and rows show the nine functional or
    anatomical ROI groups. Values are ablated minus baseline accuracy in
    percentage points; negative values therefore indicate impaired retrieval
    after ablation, whereas positive values indicate improved retrieval.
    \textbf{b,} Stimulus-identity and semantic-neighbourhood margins for each
    semantic cluster. The identity margin is the difference between matched
    brain--CLIP similarity and similarity to non-matching stimuli from the same
    semantic cluster; the semantic margin is the difference between
    same-cluster and different-cluster similarity. Markers show cluster-level
    estimates with uncertainty across participants. Hollow markers denote
    comparisons that did not survive false-discovery-rate correction
    ($q\geq0.05$); the dashed vertical line denotes zero margin.
    \textbf{c,} Functional-network composition of the most positively
    cluster-selective ROIs for all 40 semantic clusters. Radar plots show the
    distribution of up to the top 20 positively selective ROIs across the nine
    functional or anatomical groups; reference markers indicate the expected
    contribution based on each network's representation among the 379 ROIs.
    Cluster colours and ordering correspond to the retrieval-ranked semantic
    organization shown in Supplementary Fig.~\ref{fig:supp_all_clusters}.
    }
    \label{fig:supp_cluster_networks}
\end{figure}

We next extended the geometric and neuroanatomical analyses from the ten
representative clusters used in the main text to the complete set of 40 semantic
clusters (Supplementary Fig.~\ref{fig:supp_cluster_networks}). Network ablation
effects were strongly content dependent: ventral visual/object and early visual
regions produced some of the largest retrieval impairments across the semantic
space, but the magnitude and relative contribution of individual networks varied
considerably among clusters (Supplementary Fig.~\ref{fig:supp_cluster_networks}a).

The two semantic-similarity margins likewise varied across categories
(Supplementary Fig.~\ref{fig:supp_cluster_networks}b). Most clusters showed
positive stimulus-identity and/or semantic margins, although the strength and
statistical support of these effects differed across the stimulus taxonomy.
Finally, the functional-network composition of positively cluster-selective ROIs
showed marked variation across all 40 clusters
(Supplementary Fig.~\ref{fig:supp_cluster_networks}c). Together, these analyses
show that the category-dependent geometric and cortical effects highlighted in
the main text form part of a broader, continuously varying organization across
the full semantic stimulus space.

\subsection*{Expanded participant-level comparison with Ridge regression}

The main-text baseline analysis was restricted to the eight participants from BOLD5000 and NSD for whom matched results were available for \model, Ridge regression, and the MindEye-style decoder. To assess performance over a broader set of participants, we additionally compared \model with Ridge regression across all 63 dataset--participant entries for which both methods were evaluated, spanning the seven retrieval datasets (Table~\ref{tab:baselines-expanded}). Across these entries, \model achieved a higher overall mean Top-10 accuracy than Ridge regression ($22.2\%$ versus $13.1\%$) and a higher dataset-level mean in six of the seven datasets. NOD was the exception, where Ridge regression achieved the higher dataset-level mean ($12.8\%$ versus $8.1\%$). These results extend the main-text comparison by showing that the relative advantage of \model over Ridge regression is observed across a substantially broader range of datasets and participants.

\clearpage

\definecolor{headbg}{gray}{0.92}
\definecolor{meanbg}{gray}{0.96}
\definecolor{bestbg}{gray}{0.88}

{\small
\renewcommand{\arraystretch}{1.05}
\setlength{\tabcolsep}{4.5pt}
\setlength{\LTleft}{\fill}
\setlength{\LTright}{\fill}
\setlength{\LTcapwidth}{0.82\textwidth}

\begin{longtable}{llcc}
\caption{\textbf{Participant-level unique-stimulus Top-10 retrieval accuracy.}}
\label{tab:baselines-expanded}\\
\toprule
\rowcolor{headbg}
\textbf{Dataset} & \textbf{Participant} &
\textbf{Ridge regression} & \textbf{\model} \\
\midrule
\endfirsthead

\toprule
\rowcolor{headbg}
\textbf{Dataset} & \textbf{Participant} &
\textbf{Ridge regression} & \textbf{\model} \\
\midrule
\endhead

\midrule

\bottomrule
\endlastfoot

BMD & sub-01 & 19.1 & \textbf{44.9} \\
BMD & sub-02 & 17.2 & \textbf{37.5} \\
BMD & sub-03 & 14.9 & \textbf{35.2} \\
BMD & sub-04 & 17.5 & \textbf{40.7} \\
BMD & sub-05 & 13.2 & \textbf{32.4} \\
BMD & sub-06 & 19.2 & \textbf{42.8} \\
BMD & sub-07 & 11.4 & \textbf{33.5} \\
BMD & sub-08 & 12.3 & \textbf{32.5} \\
BMD & sub-09 & 16.9 & \textbf{41.3} \\
BMD & sub-10 & 17.4 & \textbf{41.0} \\
\rowcolor{meanbg}
 & Mean $\pm$ s.d. & 15.9 $\pm$ 2.8 & \textbf{38.2 $\pm$ 4.6} \\
\midrule

BOLD5000 & sub-01 & 26.8 & \textbf{46.3} \\
BOLD5000 & sub-02 & 22.6 & \textbf{38.8} \\
BOLD5000 & sub-03 & 23.3 & \textbf{42.6} \\
BOLD5000 & sub-04 & 35.9 & \textbf{41.6} \\
\rowcolor{meanbg}
 & Mean $\pm$ s.d. & 27.2 $\pm$ 6.1 & \textbf{42.3 $\pm$ 3.1} \\
\midrule

NSD & sub-01 & 18.6 & \textbf{32.8} \\
NSD & sub-02 & 17.1 & \textbf{31.1} \\
NSD & sub-03 & 15.4 & \textbf{27.3} \\
NSD & sub-04 & 17.9 & \textbf{29.1} \\
NSD & sub-05 & 20.3 & \textbf{35.9} \\
NSD & sub-06 & 23.1 & \textbf{30.1} \\
NSD & sub-07 & 3.9 & \textbf{10.6} \\
NSD & sub-08 & 13.5 & \textbf{21.6} \\
\rowcolor{meanbg}
 & Mean $\pm$ s.d. & 16.2 $\pm$ 5.8 & \textbf{27.3 $\pm$ 7.9} \\
\midrule

GOD & sub-01 & 2.2 & \textbf{42.0} \\
GOD & sub-02 & 2.9 & \textbf{45.1} \\
GOD & sub-03 & 4.7 & \textbf{52.8} \\
GOD & sub-04 & 3.8 & \textbf{56.1} \\
GOD & sub-05 & 3.4 & \textbf{45.6} \\
\rowcolor{meanbg}
 & Mean $\pm$ s.d. & 3.4 $\pm$ 1.0 & \textbf{48.3 $\pm$ 5.9} \\
\midrule

Deeprecon & sub-01 & 3.0 & \textbf{26.1} \\
Deeprecon & sub-02 & 4.8 & \textbf{28.9} \\
Deeprecon & sub-03 & 4.4 & \textbf{31.6} \\
\rowcolor{meanbg}
 & Mean $\pm$ s.d. & 4.1 $\pm$ 0.9 & \textbf{28.9 $\pm$ 2.7} \\
\midrule

THINGS & sub-01 & 7.5 & \textbf{25.2} \\
THINGS & sub-02 & 3.4 & \textbf{17.5} \\
THINGS & sub-03 & 3.0 & \textbf{18.7} \\
\rowcolor{meanbg}
 & Mean $\pm$ s.d. & 4.6 $\pm$ 2.5 & \textbf{20.4 $\pm$ 4.1} \\
\midrule

NOD & sub-01 & 1.7 & \textbf{2.1} \\
NOD & sub-02 & \textbf{4.5} & 3.3 \\
NOD & sub-03 & \textbf{8.8} & 3.8 \\
NOD & sub-04 & \textbf{4.1} & 3.2 \\
NOD & sub-05 & 3.3 & \textbf{3.6} \\
NOD & sub-06 & \textbf{5.1} & 3.1 \\
NOD & sub-07 & \textbf{4.4} & 3.3 \\
NOD & sub-08 & \textbf{5.9} & 5.2 \\
NOD & sub-09 & \textbf{4.3} & 3.1 \\
NOD & sub-10 & \textbf{4.5} & 3.5 \\
NOD & sub-11 & 9.0 & \textbf{10.0} \\
NOD & sub-12 & \textbf{14.5} & 10.0 \\
NOD & sub-13 & \textbf{16.0} & 13.5 \\
NOD & sub-14 & \textbf{25.0} & 14.0 \\
NOD & sub-15 & \textbf{16.5} & 8.5 \\
NOD & sub-16 & \textbf{13.5} & 8.0 \\
NOD & sub-17 & \textbf{17.0} & 8.0 \\
NOD & sub-18 & \textbf{16.5} & 9.0 \\
NOD & sub-19 & \textbf{12.0} & 8.0 \\
NOD & sub-20 & \textbf{24.5} & 11.0 \\
NOD & sub-21 & \textbf{21.5} & 17.5 \\
NOD & sub-22 & \textbf{27.5} & 16.0 \\
NOD & sub-23 & \textbf{10.5} & 8.0 \\
NOD & sub-24 & \textbf{14.5} & 10.0 \\
NOD & sub-25 & \textbf{23.0} & 12.5 \\
NOD & sub-26 & \textbf{8.0} & 7.0 \\
NOD & sub-27 & \textbf{21.5} & 8.0 \\
NOD & sub-28 & \textbf{20.0} & 10.5 \\
NOD & sub-29 & \textbf{19.0} & 9.5 \\
NOD & sub-30 & \textbf{8.5} & \textbf{8.5} \\
\rowcolor{meanbg}
 & Mean $\pm$ s.d. & \textbf{12.8 $\pm$ 7.6} & 8.1 $\pm$ 4.1 \\
\midrule

\rowcolor{bestbg}
Overall & Mean $\pm$ s.d. & 13.1 $\pm$ 8.0 &
\textbf{22.2 $\pm$ 15.7} \\

\end{longtable}

\begin{center}
\begin{minipage}{0.82\textwidth}
\raggedright\footnotesize
Values are percentages for the 63 dataset--participant entries evaluated
by both methods. 
\end{minipage}
\end{center}
}

\subsection*{Robustness to checkpoint selection}
\label{checkpoint}

\begin{figure}
    \centering
    \includegraphics[width=\linewidth]{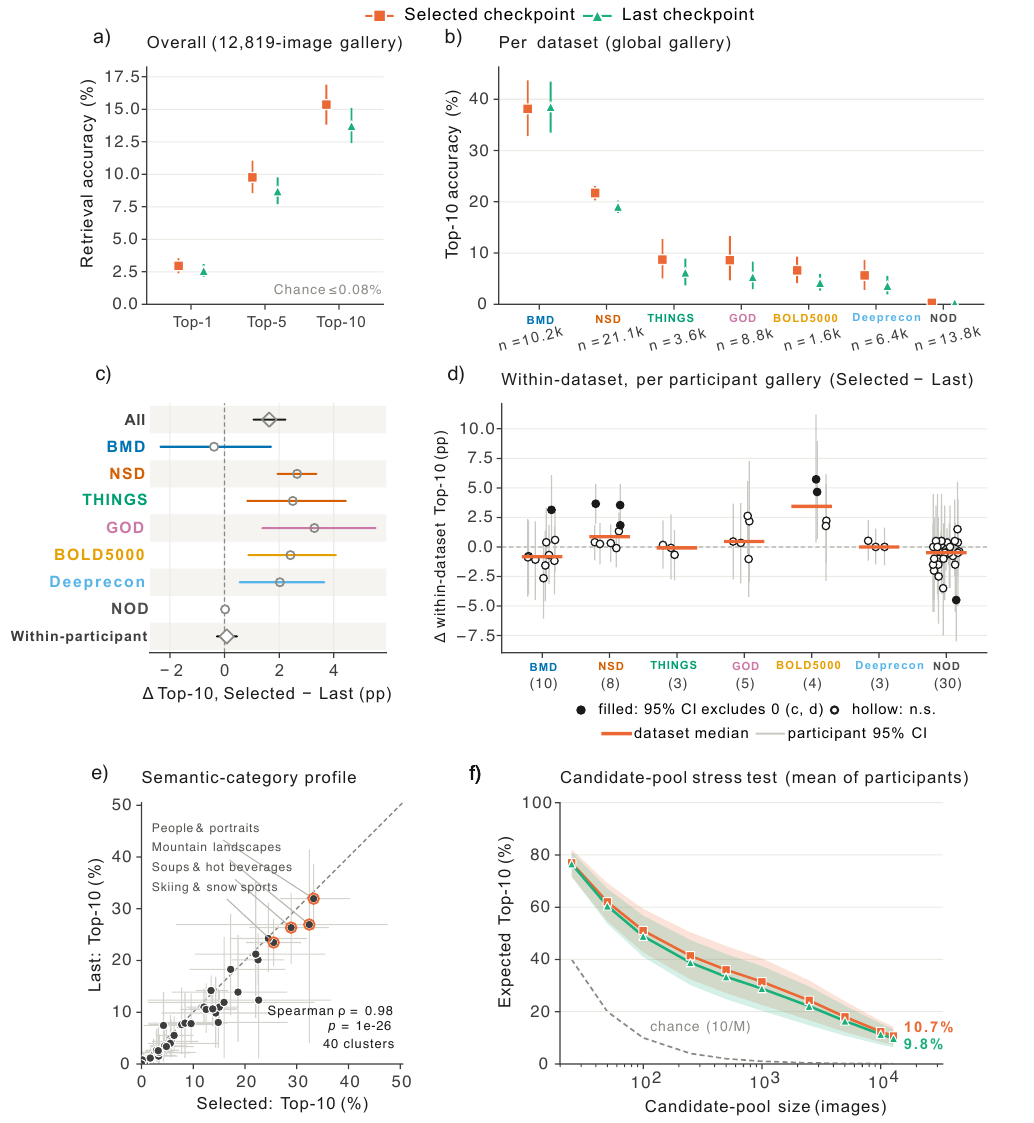}
    \caption{
    \captiontitle{Robustness of retrieval results to checkpoint selection.}
    \textbf{a,} Overall Top-1, Top-5, and Top-10 retrieval against the full
    gallery of 12,819 images. Chance Top-10 is $0.08\%$.
    \textbf{b,} Dataset-level Top-10 accuracy, with every query ranked against
    the same global gallery. $n$ denotes the number of held-out trials per
    dataset.
    \textbf{c,} Paired difference in Top-10 accuracy (Selected -- Last, in
    percentage points) overall, for each dataset, and for participant-specific
    galleries averaged across participants. Diamonds mark the two pooled
    summaries.
    \textbf{d,} Participant-level difference in Top-10 retrieval using each
    participant's own held-out gallery. Each point represents one participant;
    grey lines show 95\% CIs obtained by resampling that participant's images.
    Filled points indicate participants whose CI excludes zero, orange bars
    show dataset medians, and numbers in parentheses give the number of
    participants per dataset.
    \textbf{e,} Top-10 accuracy for each of the 40 CLIP semantic clusters at
    the selected versus last checkpoint, with 95\% CIs on both axes. The dashed
    line denotes identity, and the four best-retrieved clusters are labelled
    for orientation. Cluster ranking was highly preserved between checkpoints
    (Spearman $\rho=0.98$, $p\approx10^{-26}$).
    \textbf{f,} Candidate-pool stress test, as in
    Fig.~\ref{fig:generalization}. Expected Top-10 accuracy is shown when the
    gallery is a random subset of $M$ images that always contains the target.
    Values are computed exactly from each query's global rank using the
    hypergeometric distribution, averaged within participants and then across
    participants. Bands show 95\% CIs from participant resampling, the dashed
    line denotes chance ($10/M$), and full-gallery values are shown at right.
    The $p$-values and CIs are not corrected for multiple comparisons.
    }
    \label{fig:epoch_comp}
\end{figure}

To assess whether the main findings depended on checkpoint selection, we compared the checkpoint selected during training by validation retrieval with the final checkpoint from the same training run. Both were evaluated on the
same held-out set of 65,395 trials corresponding to 12,819 unique images from 63 participants across seven datasets (Fig.~\ref{fig:epoch_comp}). The selected
checkpoint achieved moderately higher global retrieval
(Fig.~\ref{fig:epoch_comp}a), with a $+1.6$ percentage-point difference in overall Top-10 accuracy (95\% CI, $1.1$--$2.2$ pp;
Fig.~\ref{fig:epoch_comp}c). Dataset-level performance showed a similar ordering across checkpoints (Fig.~\ref{fig:epoch_comp}b,c), whereas retrieval
using participant-specific galleries was essentially unchanged overall ($+0.1$ pp; 95\% CI, $-0.3$--$0.5$ pp;
Fig.~\ref{fig:epoch_comp}c,d). The semantic-category retrieval profile was also highly preserved (Spearman $\rho=0.98$;
Fig.~\ref{fig:epoch_comp}e), as was the decline in expected Top-10 accuracy with increasing candidate-pool size (Fig.~\ref{fig:epoch_comp}f). Thus, checkpoint selection modestly affected absolute global retrieval but did not materially alter the dataset-level, semantic, or retrieval scaling patterns
reported in the main analyses.

The selected checkpoint was chosen during training by validation Top-5 retrieval and was used for all main-text analyses; the last checkpoint corresponds to the final epoch of the same run. Accuracy is unique-stimulus Top-10 unless stated otherwise. Confidence intervals were estimated from 2,000 bootstrap resamples. For pooled retrieval analyses, resampling was performed
by image so that repeated trials of the same image remained grouped, with identical resamples used for both checkpoints to preserve pairing.

\end{document}